\documentclass[
 reprint,
 superscriptaddress,
 amsmath,amssymb,
 aps,
]{revtex4-2}

\usepackage{graphicx}
\graphicspath{{Figures/}}
\usepackage{dcolumn}
\usepackage{bm}
\usepackage{hyperref}
\usepackage{amsfonts}
\hypersetup{
    colorlinks=true,
    citecolor=blue,
    urlcolor=blue,
}
\usepackage{physics}
\usepackage{url}
\usepackage{float}
\usepackage{xifthen}
\usepackage{xcolor}
\usepackage{bm}
\usepackage{bbm}
\usepackage{soul}

\renewcommand{\autoref}[2][]{Fig.~\hyperref[#2]{\ref*{#2}\ifthenelse{\isempty{#1}}{}{(}#1\ifthenelse{\isempty{#1}}{}{)}}}

\begin{document}

\preprint{APS/123-QED}

\title{Kinemon: inductively shunted transmon artificial atom}

\author{Daria Kalacheva}
    \email{d.kalacheva@gmail.com}
 \affiliation{
 Skolkovo Institute of Science and Technology, Skolkovo Innovation Center, Moscow 121205, Russia
 }
 \affiliation{
 Moscow Institute of Physics and Technology, 9
 Institutsky Lane, Dolgoprudny 141700, Russia
 }
 \affiliation{
 National University of Science and Technology MISIS, 
 119049 Moscow, Russia
 }

\author{Gleb Fedorov}
 \affiliation{
 Moscow Institute of Physics and Technology, 9
 Institutsky Lane, Dolgoprudny 141700, Russia
 }
 \affiliation{
 National University of Science and Technology MISIS, 
 119049 Moscow, Russia
 }
 \affiliation{
 Russian Quantum Center, National University of Science and Technology MISIS, 119049 Moscow, Russia
 }
 
\author{Julia Zotova}
 \affiliation{
 Skolkovo Institute of Science and Technology, 
 Skolkovo Innovation Center, Moscow 121205, Russia
 }
 \affiliation{
 Moscow Institute of Physics and Technology, 9
 Institutsky Lane, Dolgoprudny 141700, Russia
 }
 \affiliation{
 National University of Science and Technology MISIS, 
 119049 Moscow, Russia
 }

\author{Shamil Kadyrmetov}
 \affiliation{
 Moscow Institute of Physics and Technology, 9
 Institutsky Lane, Dolgoprudny 141700, Russia
 }

\author{Alexey Kirkovskii}
 \affiliation{
 Moscow Institute of Physics and Technology, 9
 Institutsky Lane, Dolgoprudny 141700, Russia
 }

\author{Aleksei Dmitriev}
 \affiliation{
 Moscow Institute of Physics and Technology, 9
 Institutsky Lane, Dolgoprudny 141700, Russia
 }
 
\author{Oleg Astafiev}	
\affiliation{
 Skolkovo Institute of Science and Technology, Skolkovo Innovation Center, Moscow 121205, Russia
 }
 \affiliation{
 Moscow Institute of Physics and Technology, 9
 Institutsky Lane, Dolgoprudny 141700, Russia
 }
 
\date{\today}

\begin{abstract}

We experimentally investigate inductively shunted transmon-type artificial atoms as an alternative to address the challenges of low anharmonicity and the need for strong charge dispersion in superconducting quantum systems. We characterize several devices with varying geometries and parameters (Josephson energies and capacitances), and find a good agreement with calculations. Our approach allows us to retain the benefits of transmon qubit engineering and fabrication technology and high coherence, while potentially increasing anharmonicity. The approach offers an alternative platform for the development of scalable multi-qubit systems in quantum computing.

\end{abstract}


\maketitle


Superconducting artificial quantum systems, such as the capacitively shunted charge qubits (transmons and X-mons) are now commonly used to build prototypes of quantum processors because of their simple design and low decoherence rates \cite{koch2007, schreier2008, krantz2019, burnett2019}. However, scaling up quantum registers composed of low anharmonicity physical qubits faces challenges due to the uncontrolled transitions to upper states and limitations in speed of quantum operations \cite{schutjens2013, vesterinen2014, chen2016, mcewen2021, bultink2020, miao2022}. Additionally, non-negligible charge dispersion of the higher energy levels complicates the use of such artificial atoms as qudits \cite{roy2022}. These problems drive the search for alternative physical qubits and materials \cite{pechenezhskiy2020, gyenis2021, hyyppa2022, hassani2022}. 

While retaining simplicity in fabrication and operation, together with charge noise insensitivity, one can increase the non-linearity of a transmon by decreasing its shunting capacitance and, at the same time, shunting it by a linear inductance. Strictly speaking, this modification produces a flux qubit \cite{orlando1999, chiorescu2003, you2007, yan2016, peltonen2018}, more specifically, an rf-SQUID or a fluxonium \cite{moskalenko2022}; however, its parameters can be chosen so that the resulting eigenstates are transmon-like, living in a single-well potential, not a two-well one. The latter helps to avoid the exponential sensitivity of the transition frequencies to the Josephson energy variations, which has so far limited the applications of flux qubits in multi-qubit devices. Also, one can expect that the inductive shunt will remove charge dispersion for arbitrarily high energy states. 

In this study, we explore a new hybrid design combining the transmon circuit with a compact kinetic inductor -- a kinemon (kinetic-inductance-shunted transmon) artificial atom. We design and investigate experimentally a family of such systems with various combinations of Josephson energy $E_J$, inductive energy $E_L$ and charging energy $E_C$ \cite{krantz2019}. We also show that the inductive element can be placed inside an $\alpha$-SQUID \cite{paauw2009} which, for a correct ratio of resulting loop areas, opens a way to modulate the effective Josephson energy, while keeping the parabolic potential contribution fixed. Importantly, we find that, confirming our previous tests of coplanar resonators \cite{kalacheva2023}, the kinetic inductor based on aluminum ultra-thin-film does not cause any noticeable deterioration of the coherence times. Finally, as this kind of inductor exhibits relatively good reproducibility in fabrication, we find it a promising component for future quantum circuits. 




\begin{figure*}[htp]
\includegraphics[width=1\linewidth]{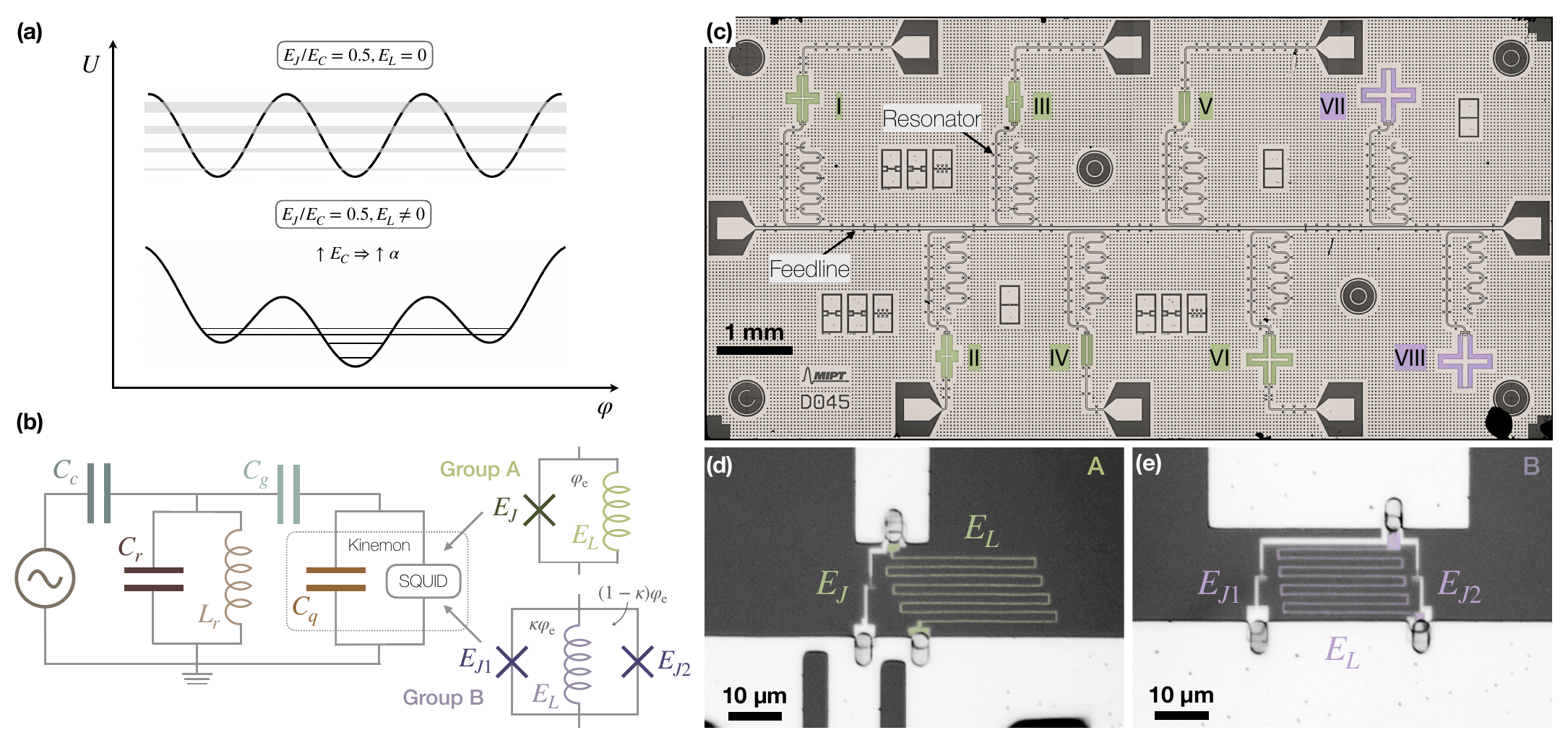}
\caption{\label{fig:sample} \textbf{(a)} The potential (solid black line) and the energy levels (gray lines) for the conventional transmon (top) and for the inductivly shunted one (bottom) for $\varphi_\mathrm{e} = 0$. \textbf{(b)} Circuit model of the sample. SQUIDs consists of Josephson junction and a kinetic wire with the energies $E_J$ and $E_L$ respectively and are shunted by a large capacitance with an energy $E_C$ forming a kinemon atom. Group A (green color) represents assymetric kinemons with different energy ratios (Kinemons I - VI). Group B (violet color) represents symmetric kinemons with two JJs and a kinetic inductance wire between them (Kinemons VII - VIII). \textbf{(c)} Overview optical image of the fabricated sample. Microfabrication design includes two group of art. atoms with different topologies and variations of $E_C$, $E_L$ and $E_J$. \textbf{(d,\,e)} Enlarged optical image of two different kinemon modification.} 
\end{figure*}

Without the inductive shunt, a Josephson junction is characterized by a periodic potential $U_J = -E_J\cos\varphi$, where $\varphi$ is the phase across the junction [Fig. \ref{fig:sample}(a)]. Adding a small parallel capacitance to the circuit results in the formation of energy bands of Bloch waves in the periodic potential \cite{likharev1985}, which can be represented as $\psi'(\varphi) = e^{i \frac{q'}{2e} \varphi} u(\varphi)$, so that $\psi'(\varphi) \neq \psi'(\varphi+2\pi)$ and only $u(\varphi) = u(\varphi + 2\pi)$.  Here, $q'$ represents the quasicharge, being the analogue of the crystal momentum in solids, and the energies inside the band are periodic functions of $q'$ [condensed matter book]. We note that for a charge qubit with discrete number of Cooper pairs allowed on the island, we can apply the rotor analogy \cite{koch2007, devoret2021does}, so that the states after a full rotation are indistinguishable. Then, the wave function is $2\pi$-periodic in the $\varphi$ representation, and the bands seem to disappear. However, as there is a mathematical correspondence between the quasicharge $q'$ and the induced charge $n_g$: $n_g = q'/2e$, the energy configuration of the system still exhibits the same oscillatory behavior as a function of $n_g$. While a larger capacitance localizes the lower energy states in distinct potential wells and significantly reduces the widths of the lowest bands, higher lying bands remain open, as shown in Fig. \ref{fig:sample}(a), and thus are still sensitive to the induced charge. Moreover, increasing the capacitance always comes at the cost of reducing the anharmonicity $\alpha$ \cite{koch2007}.

To completely prevent the formation of energy bands (remove any energy dependence on the induced charge $n_g$), it is necessary to disrupt the periodicity. This can be achieved by implementing a shunting inductance, which introduces a parabolic potential $U_L = E_L \varphi^2/2$ to $U_J$ as exemplified in Fig.~\ref{fig:sample}(a). As a result, the wavefunctions of the lowest states become localized in the central well. Note that in the case of a small $E_L$ ($\leq E_J$), the anharmonicity and the energy structure are predominantly determined by $E_J$ and $E_C$. This is because the Josephson energy $E_J$ governs the energy landscape of the system, with the smaller inductive energy $E_L$ providing only a minor perturbation. Larger $E_L/E_J$ results in less anharmonicity but can be compensated by increasing $E_C$. In the present work, though, we study inductively shunted artificial atoms fabricated using the standard transmon technology ($E_C \ll E_J \lesssim E_L$) and aim to verify their properties and coherence.   

\begin{figure*}[htp]
\includegraphics[width=1\linewidth]{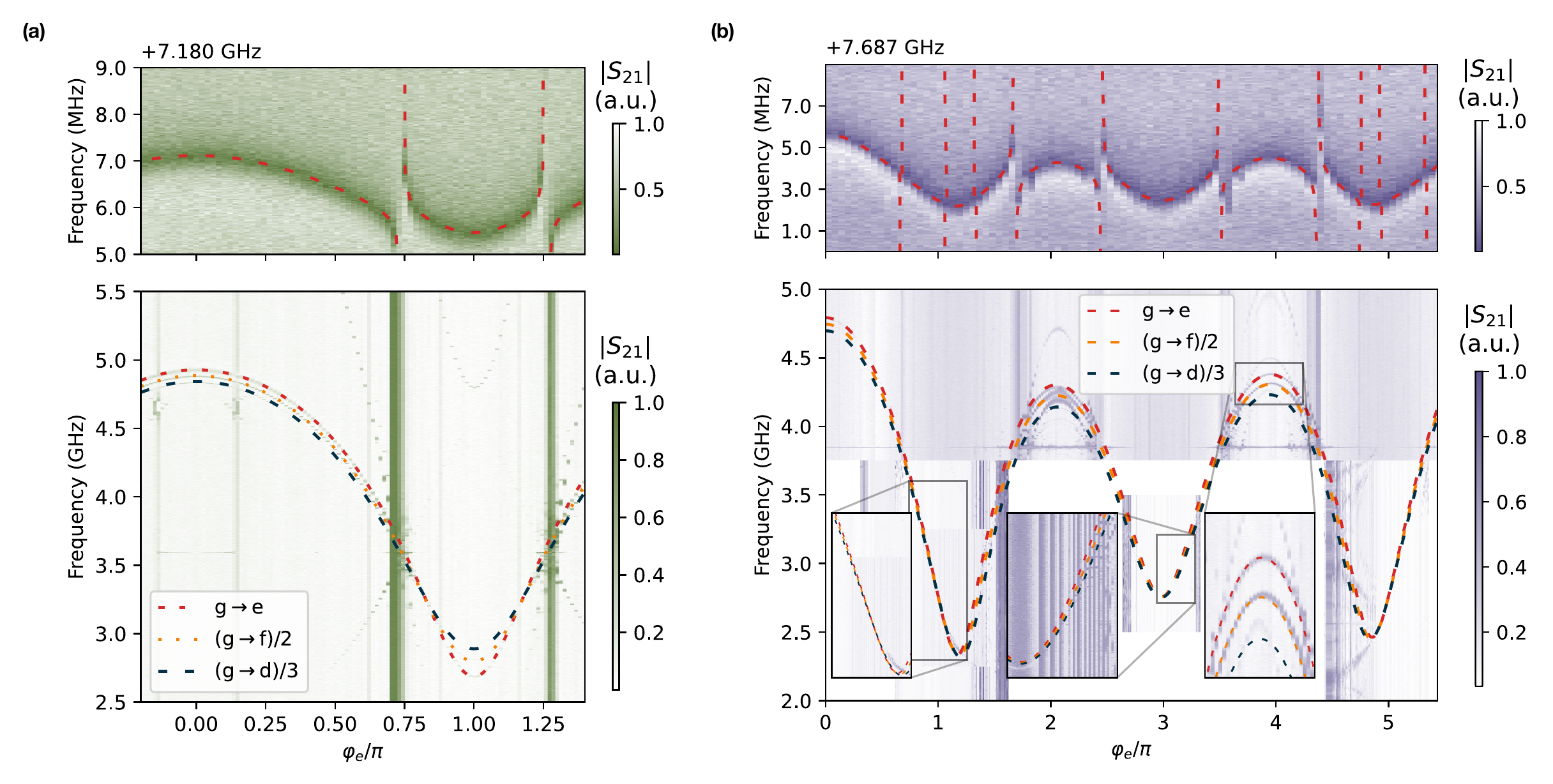}
\caption{\label{fig:spectra} \textbf{(a)} Kinemon I: \textbf{(Top)} Transmission spectroscopy of the coupled resonator showing feedline frequency response depending on the external flux bias $\varphi_\mathrm{e}$. \textbf{(Bottom)} Experimental two-tone spectroscopy, displaying the magnitude of the readout signal at a properly chosen readout frequency as a function of flux bias $\varphi_\mathrm{e}$ and kinemon atom excitation frequency. For Group A, the anharmonicity sign changes at the $\varphi_\mathrm{e} = 0.75\,\pi$ and gets back at $\varphi_\mathrm{e} = 1.25\,\pi$. Numerical simulation of the spectrum reproducing experimental results with labeled transitions. \textbf{(b)} Kinemon VII: \textbf{(Top)} Transmission spectroscopy of the coupled resonator. \textbf{(Bottom)} Experimental two-tone spectroscopy, with insets (middle and right) showing magnifications at the bottom and top flux sweet spots. For Group B, the regimes of harmonic oscillator are noticeable near $\varphi_\mathrm{e} = \pi + 2\pi k,\ k \in Z$ (left and middle inset), while at $\varphi_\mathrm{e} \approx 1.15\,\pi$ only the three lowest levels are equidistant and shown in the left inset.}
\end{figure*}

A schematic equivalent circuit of kinemon artificial atom is depicted in Fig.~\ref{fig:sample}(b), illustrating two different configurations of qubits investigated in this work. Qubits consist of two main parts: $\text{Al}/\text{Al}\text{O}_x/\text{Al}$ Josephson junctions and a kinetic inductance wire with energy $E_L = \Phi_0^2/4\pi^2 L_k$ due to the wire inductance $L_k$ to form a SQUID loop. The first circuit scheme operates with an asymmetric topology, utilizing a small single loop formed by an aluminum wire interrupted by a single Josephson junction (Group A in Fig.~\ref{fig:sample}(b)). A key characteristic of this mode is the variation of shunting capacitance values ($E_C = 2e^2/C$) to examine coherence time. The second circuit scheme employs a symmetric topology that merges the benefits of both rf-SQUID and transmon designs (Group B in Fig.~\ref{fig:sample}(b)). This approach aims to enhance qubit performance by incorporating a double-loop architecture connected by a shared kinetic inductance wire.

The common Hamiltonian, which covers both the considered geometries (single- and double-loop circuits), experesses as follows
\begin{equation}
\begin{split}
	\mathcal{\hat{H}} = -E_C \frac{\partial^2}{\partial\varphi^2} &+ \frac12 E_L\varphi^2 - E_{J1}\cos(\varphi + \kappa\varphi_\mathrm{e})\\ 
    &-E_{J2}\cos\left(\varphi - (1-\kappa)\varphi_\mathrm{e}\right),
\end{split}\label{eq:H_kinemon}
\end{equation}
where $\varphi_\mathrm{e}$ is the total flux phase induced by an external magnetic field, $\kappa$ is the coefficient of $\varphi_\mathrm{e}$ distribution between two loops, $E_{J1}$ and $E_{J2}$ are the Josephson energies of the junctions in each loop according to Fig. \ref{fig:sample}(b). The Hamiltonian for Group A is obtained by putting $E_{J2}=0$. 

An optical image of the sample is presented in Fig.~\ref{fig:sample}(c), depicting the microfabricated superconducting circuit containing eight kinemon artificial atoms. SQUIDs are tiny loops of superconducting wire and Josephson junctions connected to other circuit elements, such as capacitors and resonators. The SQUIDs are highlighted in green and violet false colors depending on their architecture Fig.~\ref{fig:sample}(d,\,e). 

The fabrication of the kinetic inductance wire is done as follows. A silicon substrate is cooled down by liquid nitrogen during metal deposition to obtain uniform films up to 3.5 nm thick. It is known that aluminum films, deposited at room temperature are negatively impacted by formation of granules. Cold film deposition allows us to fabricate long ultrathin wires with high degree of homogeneity \cite{kalacheva2023}. The detailed fabrication process described in App.~\ref{app:fabrication}. To achieve the necessary inductive energy $E_L$, 200 nm wide and 8 nm thick aluminum wires with varying lengths are integrated into the circuit. Since the kinetic inductance per square of such a film is about 0.03 nH/$\square$, the wire length in the device ranges from 80 to 240 $\mu$m, depending on the desired $E_L$.

\begin{table*}[htp]
	\caption{\label{tab:meas_params}
		Kinemons parameters extracted by fitting. Coherence times are given at $\varphi_\mathrm{e} = 0$.}
	\begin{ruledtabular}
		\begin{tabular}{ccccccccc}
			&I&II&III&IV&V&VI&VII&VIII\\
			\hline			
			$E_J/h$, $\text{GHz}$&$5.38$&$6.00$&$4.00$&$2.92$&$2.44$&$5.90$&$8.61$&$14.00$\\
						
			$E_C/h$, $\text{GHz}$&$0.90$&$1.10$&$1.50$&$1.95$&$1.80$&$0.7$&$0.47$&$0.32$\\
			
			$E_L/h$, $\text{GHz}$&$8.59$&$8.75$&$7.40$&$8.40$&$9.07$&$14.65$&$8.11$&$12.2$\\
			

			$\omega^{(t)}_{01}/2\pi$, $\text{GHz}$&$4.947$&$5.596$&$5.719$&$6.508$&$6.359$&$5.312$&$4.769$&$5.008$\\

            $\omega_r/2\pi$, $\text{GHz}$&$7.185$&$7.284$&$7.341$&$7.433$&$7.495$&$7.608$&$7.688$&$7.779$\\

            $g_s/2\pi$, $\text{MHz}$&$64$&$44$&$34$&$35$&$34$&$90$&$83$&$68$\\

            $\alpha^{(t)}/h$, $\text{MHz}$&$-86$ &$-118$&$-131$&$-116$&$-87$& $-49$&$-84$&$-80$\\
			
            $\alpha^{(b)}/h$, $\text{MHz}$&$219$ & $301$ & $257$ & $182$ & $124$ & $96$&$-$&$-$\\
			
			$T_1$, $\mu s$&$17.92 \pm 0.95$&$17.56 \pm 0.86$&$19.45 \pm 1.62$&$8.95 \pm 0.33$&$8.61 \pm 0.29$&$20.39 \pm 0.93$&$19.18 \pm 0.78$&$14.83 \pm 0.87$\\
	
			$T_2$, $\mu s$&$11.45 \pm 0.95$&$17.30 \pm 1.65$&$11.63 \pm 1.20$&$7.80 \pm 0.85$&$9.69 \pm 0.42$&$13.98 \pm 0.75$&$6.59 \pm 0.30$&$12.28 \pm 0.65$\\
			
			$T_{2E}$, $\mu s$&$7.92 \pm 2.35$&$20.84 \pm 1.11$&$-$&$8.05 \pm 0.83$&$14.87 \pm 0.56$&$18.25 \pm 1.06$&$12.73 \pm 0.50$&$13.32 \pm 1.02$\\

		\end{tabular}
	\end{ruledtabular}
\end{table*}

\begin{figure}[htp]
\includegraphics[width=1\linewidth]{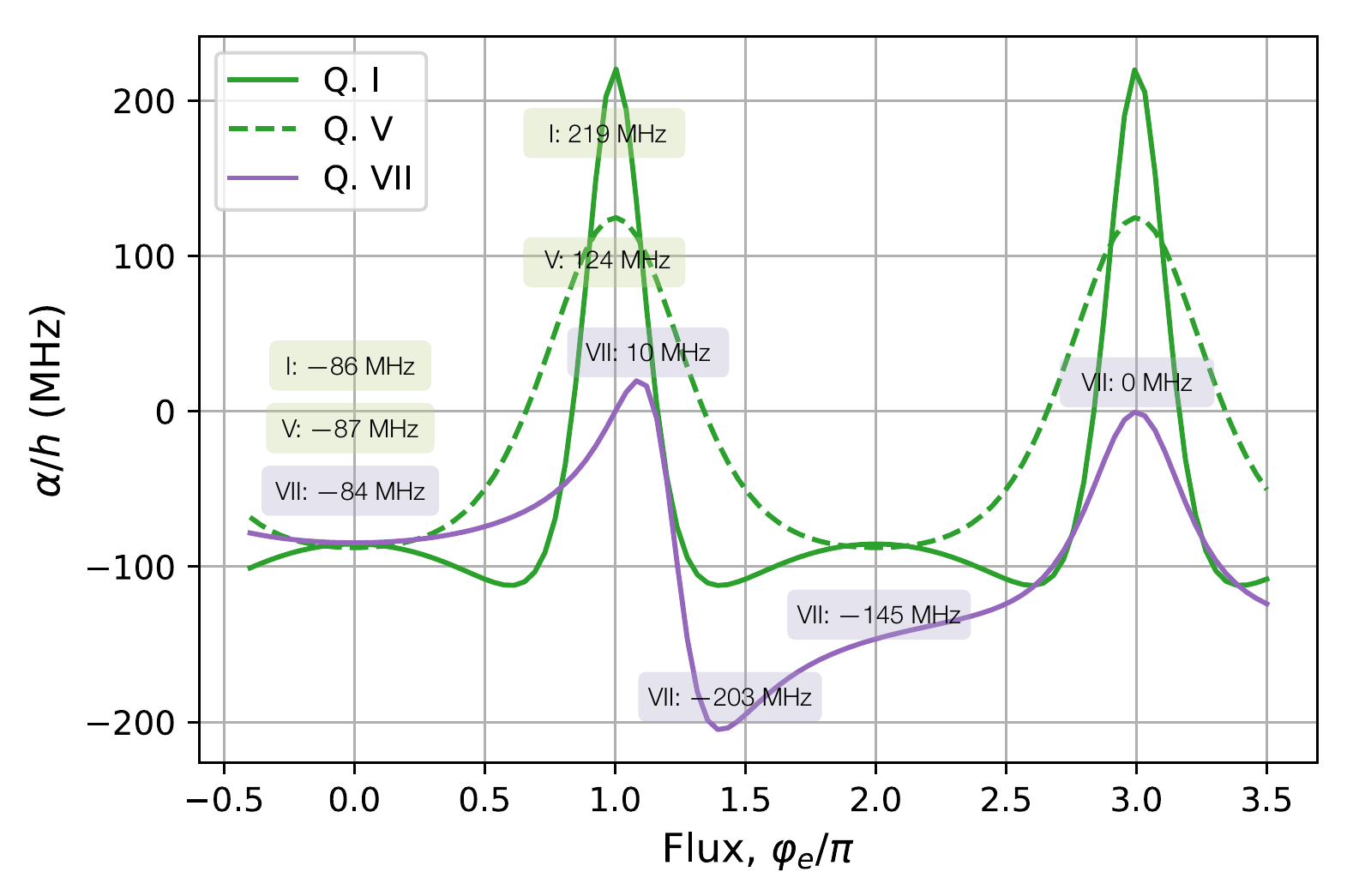}
\caption{\label{fig:alpha} Anharmonicity as a function of the external flux bias $\varphi_\mathrm{e}$ for kinemon I, V an VII. For kinemons I and V, a zero anharmonicity implies that the first three energy levels are equidistant. As for kinemon VII, the anharmonicity becomes exactly zero $\varphi_\mathrm{e} = \pi$ at  as it transitions into the harmonic regime, while at $\varphi_\mathrm{e} \approx 1.15\,\pi$ the behavior is similar to Group A.} 
\end{figure}


In Fig.~\ref{fig:spectra}(a) top, we display the data of transmission spectroscopy of the sample via the feedline, showing the microwave response of the readout resonator I as a function of the flux bias $\varphi_\mathrm{e}$ through the SQUID of kinemon I (Group A). See all resonators spectra in Fig.~\ref{fig:dsts_all} in App.~\ref{app:model}. The pattern is a combination of a smooth dependence, formed by the kinemon first excited state located below the resonator frequency, and an avoided crossing pattern with the second excited state \cite{fedorov2019}. Direct observation of these transitions is enabled by the cross-Kerr dispersive spectroscopy, for which the data are displayed in Fig.~\ref{fig:spectra}(a) bottom. The minimum frequency (one-half flux quantum) and the maximum frequency (zero flux) are considered the flux `sweet spots'. Fig.~\ref{fig:spectra}(a) bottom, also presents the fits to the experimental transition frequencies based on the circuit Hamiltonian (Eq. \ref{eq:H_kinemon}). The fits align well with the experimental transition frequencies near the sweet spots $\varphi_\mathrm{e} = 0$ and $\varphi_\mathrm{e} = \pi/2$. The single-photon transition between the ground $|0\rangle$ and excited $|1\rangle$ states occurs at $f_{01} = 4.947\,\text{GHz}$ at $\varphi_\mathrm{e} = 0$. Furthermore, the bottom sweet spot of the transition from $|0\rangle$ to $|2\rangle$ is at $f_{02} = 5.6\,\text{GHz}$ and not visible at $\varphi_\mathrm{e} = \pi$ and corresponds to anti-crossings in the transmission spectroscopy. However, the two-photon transition can be observed and is associated with the spectroscopic line at $f_{02/2} = 4.8,\text{GHz}$. Also, at $\varphi_\mathrm{e} \approx 0.75\,\pi$ and $\varphi_\mathrm{e} \approx 1.25\,\pi$ there are regimes when the first three energy levels are equidistant and are the pivot points in the anharmonicity sign. The atom with zero anharmonicity for the first two transitions \cite{bardetski2003} ($\ket{0}\to \ket{1}$ and $\ket{1}\to \ket{2}$), which is intermediate between a two-level system (TLS) and a harmonic oscillator, may find interesting applications in the generation of non-classical light \cite{gasparinetti2019}. All extracted parameters from the fits are summarised in Table \ref{tab:meas_params}. The extracted parameters are also used to make predictions and test the validity of the theoretical model employed in the analysis. The experimental setup and measurement equipment are described in App.~\ref{app:setup}.

The symmetric kinemons (Group B) are also characterized using spectroscopic measurements. The kinemon VII transmission and two-tone spectra are presented in Fig.~\ref{fig:spectra}(b), where in insets we highlight several distinct features at and near the sweet spots. In the current scheme, we operate in the regime where $E_{J1} = E_{J2}$, leading to zero anharmonicity at bottom sweet spots ($\alpha = 0$ for $\varphi_\mathrm{e} = \pi + 2\pi k,\ k \in Z$). In other words, the Josephson energy is cancelled, when a half flux quantum $\Phi_0$ penetrates through the SQUID, causing a transition into the harmonic regime. The modulation of the frequency at the top sweet spots corresponds to the non-identical areas of the SQUIDs. By fitting the spectrum, we evaluate $\kappa$ to 0.35 and 0.37 for kinemon VII and VIII, respectively, which is in a good agreement with the design areas. These values determine the locations of non-periodic spots of three-level equidistance (similar to Group A) at $\varphi_\mathrm{e} \approx 1.15\,\pi$ and $\varphi_\mathrm{e} \approx 4.9\,\pi$, respectively.

To illustrate better the behavior of the spectra, we plot the anharmonicity of the kinemons I, V and VII vs. the magnetic flux in \autoref{fig:alpha}. For all devices, we calculate the anharmonicity as $\alpha/h = 2\times(f_{02}/2 - f_{01})$. For kinemon I this yields -86 MHz and 219 MHz for the top and bottom sweet spots, respectively. Kinemon V shows a lower peak anharmonicity, but in overall demonstrates a flatter dependence on $\varphi_\mathrm{e}$ due to an increased $E_C$ value. For kinemon VII, zero anharmonicity at $\varphi = \pi + 2\pi k,\ k \in Z$ is associated with a transition into a harmonic oscillator; the additional feature at  $\varphi \approx 1.15\,\pi$ mirrors the effect observed in kinemons I and V, corresponding to the three-level equidistance. Interestingly, for this qubit the anharmonicity is mostly negative and is higher in absolute value in the top sweet spot near $\varphi_
\mathrm{e} = 2\,\pi$ than at $\varphi_
\mathrm{e} = 0$. Another peculiarity is that due to the non-trivial $\kappa$-dependence of potential one can find a sign change of anharmonicity at $\varphi = \pi + 2\pi k,\ k \in Z$. However, if the flux is simultaneously near one of the other set of special points $\varphi_\mathrm e / \pi = \frac{1 + 2 k'}{2 \kappa - 1},\  k' \in Z$ (for our case, this is near $\varphi_\mathrm e \approx 3\, \pi,\  k' = -1$), the sign of $\alpha$ remains the same.

\begin{figure*}[htp]
\includegraphics[width=1\linewidth]{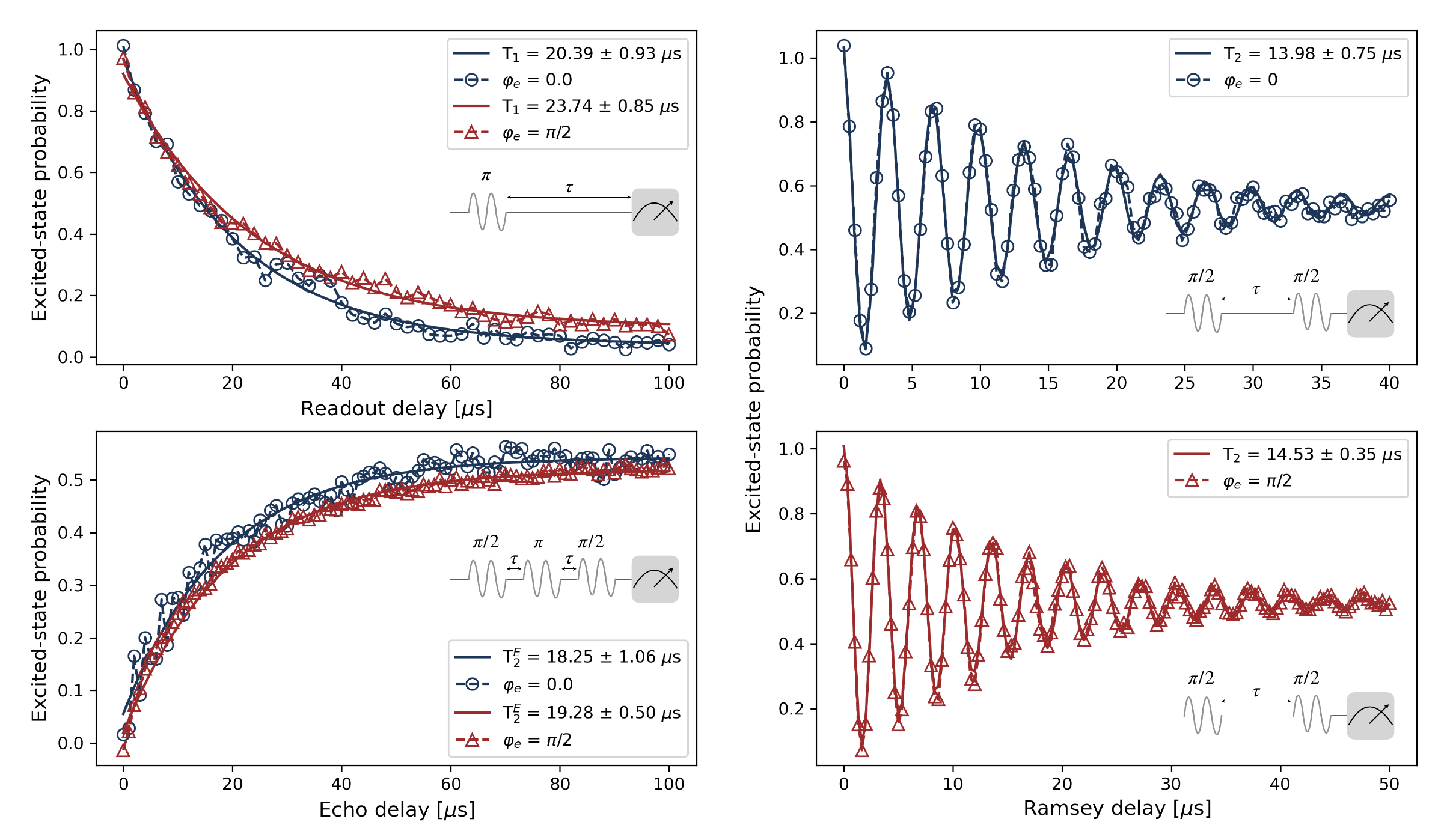}
\caption{\label{fig:qVI_times} Energy relaxation and coherence times for kinemon VI measured at different flux bias. The analysis considers two flux bias conditions, $\varphi_\mathrm{e} = 0$ (navy circles) and $\varphi_\mathrm{e} = \pi/2$ (red triangles). The top left subplot reveals the $T_1$ time at the top sweet spot, demonstrating stability with a value of $20.39\,\pm\,0.93\,\mu\text{s}$. Triangles on the red line represent higher $T_1$ values, $23.74\,\pm\,0.85\,\mu\text{s}$, suggesting slower relaxation at $\varphi_\mathrm{e} = \pi/2$. The echo coherence time ($T^E_2$) is depicted in the bottom left subplot. Measurements display times of $18.25\,\pm\,1.06\,\mu\text{s}$ and $19.28\,\pm\,0.50\,\mu\text{s}$ for $\varphi_\mathrm{e} = 0$ and $\varphi_\mathrm{e} = \pi/2$ respectively. Lastly, the remaining subplot illustrates Ramsey coherence times ($T_2$), yielding values of $13.98\,\pm\,0.75\,\mu\text{s}$ and $14.53\,\pm\,0.35\,\mu\text{s}$ at the respective sweet spots.} 
\end{figure*}

Finally, we analyse the energy relaxation time $T_1$, the Ramsey coherence time $T_2$, and the echo coherence time $T^E_2$ (See in Table \ref{tab:meas_params}). The experimental results for kinemon VI are presented in Fig.~\ref{fig:qVI_times}, which provides a comparative analysis of relaxation times under two distinct flux bias conditions, specifically at $\varphi_\mathrm{e} = 0$ (represented in navy color) and $\varphi_\mathrm{e} = \pi/2$ (represented in red color). Corresponding measurement protocols are presented on each subplot. We also measure the coherence outside the sweet spot and find the characteristic times to be about 600 ns and 200 ns for $T^E_2$ and $T_2$, respectively; the reduction is probably caused by the flux noise due to the insufficient filtering. The relaxation time remains the same with respect to the flux value. The experimental data for $T_1$ are presented in the top left subplot. At the top sweet spot, a specific value noted as $20.39\,\pm\,0.93\,\mu\text{s}$. The values represented by triangles on the red line seem slightly higher than at $\varphi_\mathrm{e} = 0$, indicating a slower relaxation rate at $\varphi_\mathrm{e} = \pi/2$ with a particular time noted as $23.74\,\pm 0.85\,\mu\text{s}$. The echo coherence time is plotted in the bottom left subplot, with measurement results of $18.25 \pm 1.06\,\mu\text{s}$ and $19.28\,\pm\,0.50\,\mu\text{s}$ at flux biases of $\varphi_\mathrm{e} = 0$ and $\varphi_\mathrm{e} = \pi/2$, respectively. The Ramsey coherence times are presented in the remaining subplot and yield $13.98\,\pm\,0.75\,\mu\text{s}$ and $14.53\,\pm\,0.35\,\mu\text{s}$ at both sweet spots, respectively. Also, additional measurements of conventional transmons fabricated with the same technological process give about $14\,\mu\text{s}$, $8\,\mu\text{s}$ and $9\,\mu\text{s}$ for $T_1$, $T_2$, $T^E_2$, respectively; however, observed performance improvement of kinemons could be caused by differences between the measurement setups.


The use of low-loss material, such as ultra-thin aluminum film inductors, not only improves device scalability but also enhances performance compared to other materials \cite{peltonen2018, winkel2020, rieger2023}. These observations indicate that the shunting capacitance may be further reduced to make the kinemon design even more compact compared to the transmons. For example, the typical value of $E_c = 1.5$\,\text{GHz} for transmons \cite{stehlik2021, googleAI2023} could be increased up to 5-6\,GHz for kinemons which will scale down the capacitors, which is promising for the scalability of multi-qubit systems. Additionally, the advantage of having a sign-changing anharmonicity attracts interest in waveguide quantum optics \cite{gu2017}, for example, allowing to emit pairs of correlated photons \cite{gasparinetti2017} and observe nonlinear intermodulation processes \cite{honigl2018} or could help to optimise gate errors caused by a parasitic partial CPHASE operation induced by high-order coupling \cite{yan2018, zhao2020}, or be useful for new regimes of Bose-Hubbard model simulators \cite{fedorov2021, zhang2023}.


In conclusion, this study demonstrates that inductively shunted transmon qubits, utilizing ultra-thin aluminum film inductors, provide a promising platform for scalable quantum computing applications. To increase anharmonicity and strike a balance between high anharmonicity and low charge noise sensitivity, a reduction in capacitance should be considered. Thus, introducing a non-zero kinetic inductance component can address this issue.  While this study represents a significant step towards realizing large-scale, practical quantum computing systems, further research is required to fully explore and validate the potential of this approach in the field of quantum computing. 

\section*{Data Availability}

The raw data that support the findings of this study are available on a reasonable request from the corresponding author.

\begin{acknowledgments}
The authors are grateful to Russian Science Foundation Project Grant No.\,21-42-00025 for financial support. The sample was fabricated using equipment of MIPT Shared Facilities Center.  
\end{acknowledgments}

\appendix

\section{Design of the qubit samples}\label{app:design}

A layer-by-layer design was generated using the Klayout-python library \cite{klayoutSh}, which automates the design of superconducting quantum circuits. This library utilizes the KLayout layout design program API and enables the execution of arbitrary Python code through an embedded interpreter. The library specializes in designing microwave and superconducting qubit planar designs, including drawing patterns, simulation, and domain-specific design rule checkers.

\section{Device fabrication}\label{app:fabrication}

The device fabrication steps consist of five main stages: ground plane construction, nanofabrication of Josephson junction and kinetic wire, bandage deposition, and air-bridges construction.

We start with silicon substrate treatment, which includes piranha etching and BHF dipping \cite{bruno2015, kalacheva2020}. The substrate is then immediately placed in the Plassys e-beam evaporation system, and a 100 nm 99.999\% aluminum film is evaporated. The metallized substrate is spin-coated with optical resist AZ1517. The coplanar waveguide feedline, resonators, qubit capacitors, and ground plane hole array are patterned using a laser maskless optical lithography system, followed by dry etching of the optical resist mask structure in BCl3/Cl2 inductively coupled plasma. Residual resist is then removed in N-methyl-2-pyrrolidone (NMP) and cleaned in O2 plasma.

The next step includes hard-mask preparation \cite{liechao2012}. The substrate is spin-coated with polymer resist PMGI SF9. Then, a 30 nm tungsten nanolayer is deposited in a Torr magnetron sputtering system, followed by ARP-04 resist coating. Josephson junctions are patterned by electron lithography and evaporated using the Dolan bridge technique \cite{dolan1977}, followed by lift-off in NMP. To form the tunnel barrier, the first 25 nm aluminum junction electrode is oxidized at 40 mBar. Then, a 45 nm electrode is evaporated and preventively oxidized at 10 mBar. Residual resist is removed in NMP and cleaned in O2 plasma.

The kinetic part is fabricated during an additional cycle of e-beam lithography. We use a single layer of ARP-04 e-beam resist to construct the pattern. After development, an 8 nm aluminum film is evaporated at the Plassys stage temperature of 170 K at a normal angle. Residual resist is then removed in NMP and cleaned in O2 plasma.

Good galvanic contact between the layers is obtained by aluminum bandages \cite{osman2021}. We use a similar process of single layer mask as for the kinetic part above, but without cooling. A 150 nm aluminum film is evaporated with in-situ Ar ion milling. Residual resist is removed during the lift-off process in NMP and cleaned in O2 plasma.

Due to the presence of coplanar lines on the ground plane, we need to achieve a uniform electrical potential; therefore, the final stage of sample fabrication is the implementation of aluminum free-standing air-bridges \cite{chen2014}. A 7 µm layer of SPR220 photoresist is spin-coated, and the base layer is patterned using a laser maskless optical lithography system. After development, the substrate is heated to create a height gradient on the resist edges, followed by a 600 nm aluminum evaporation with in-situ Ar ion etching. A second layer of SPR220 photoresist is used to form a bridge structure. Finally, the excess metal is dry-etched in BCl3/Cl2 inductively coupled plasma. Residual resist is then removed in NMP and cleaned in O2 plasma.

\begin{figure}
\includegraphics[width=1\linewidth]{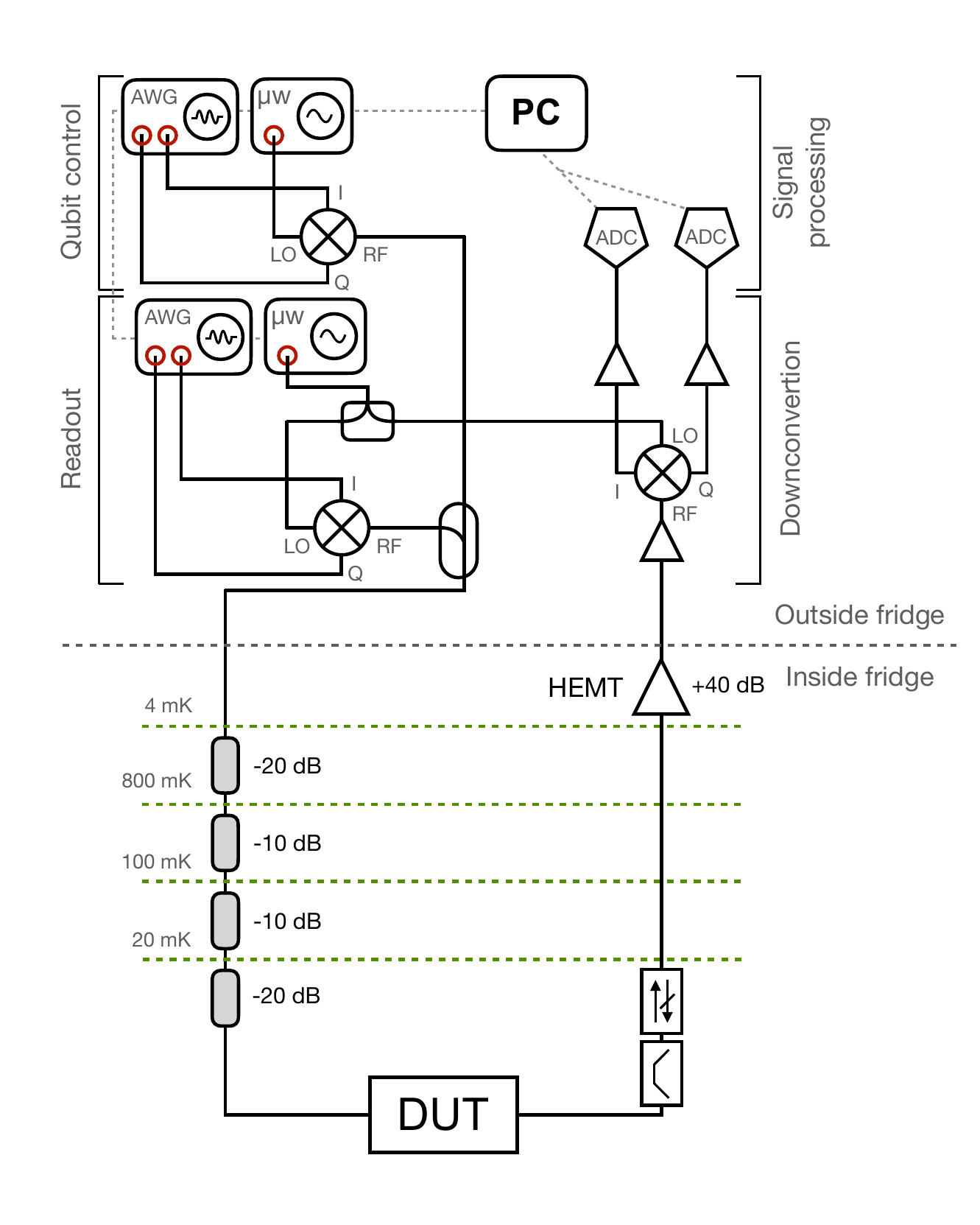}
\caption{\label{fig:setup} The schematic of experimental setup to measure the sample depicting room temperature equipment and line configuration inside the BlueFors dilution refrigerator, with the base temperature of 10 mK.} 
\end{figure}

\section{Microwave experimental setup}\label{app:setup}

\begin{figure}[h!]
\includegraphics[width=1\linewidth]{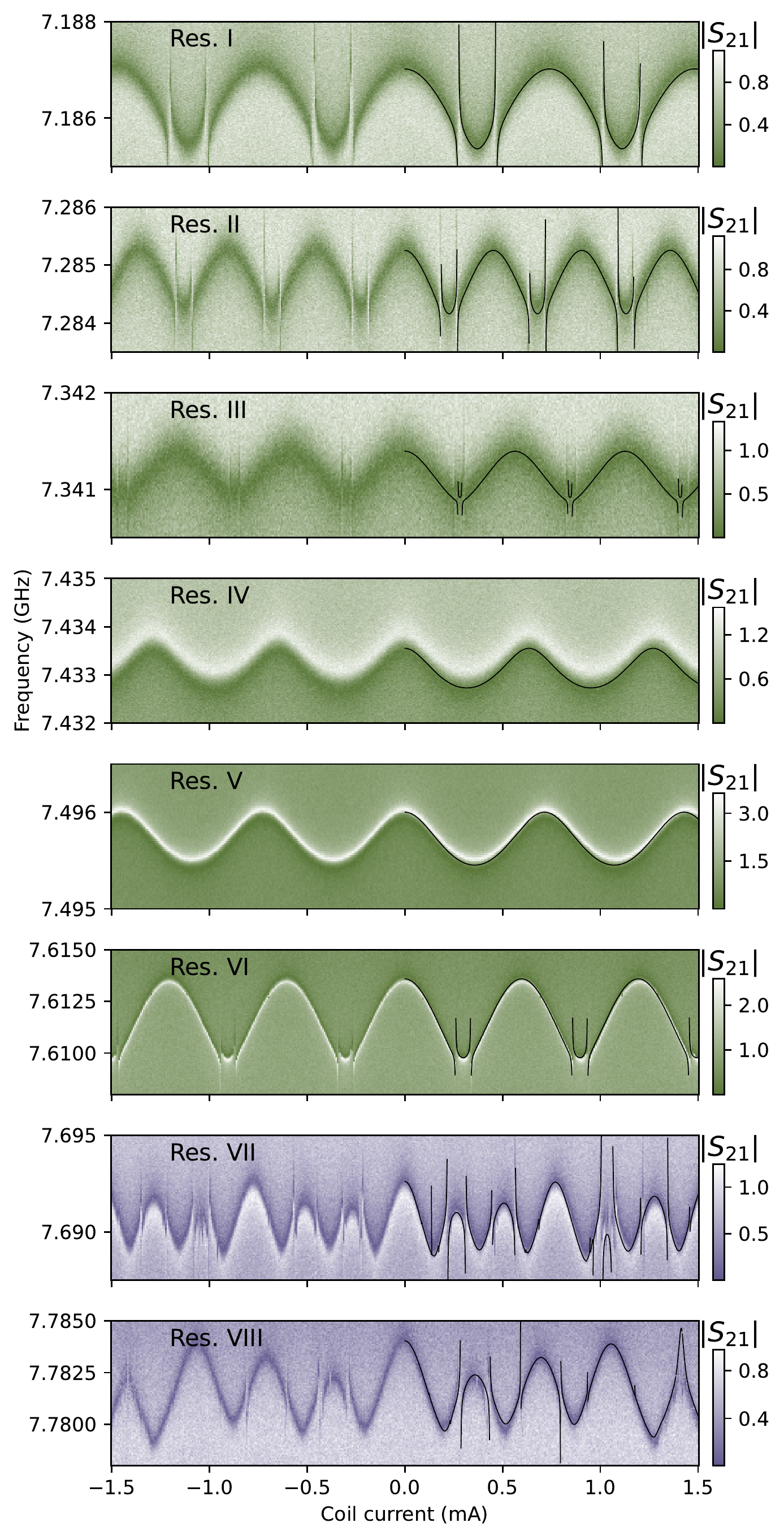}
\caption{\label{fig:dsts_all} Transmission spectroscopy for all resonators coupled to kinemons with varying energy ratios. $|S_{21}|$ includes the attenuation and amplification in the measurement chain. Avoided crossings occur when a qubit's $\ket{0} \to \ket{2}$ transition (or even higher levels for kinemons VII and VIII) intersects with its readout cavity frequency. Some very sharp avoided crossings are just barely resolved in the simulations. Also, some of the predicted features are smeared in experiment due to the power broadening effects (located around 1.1 mA for VII and 1.4 mA for VIII).}
\end{figure}

The device under investigation is measured in a dilution refrigerator at 10 mK (\autoref{fig:setup}). Signals are generated using an arbitrary waveform generator (Keysight M3202A) and RF-synthesizers (SignalCore 5502A), followed by upconversion in IQ-mixers (Marki IQ4509 and IQ0307). Excitation and dispersive readout signals are combined using a directional coupler and sent to the fridge, where they are attenuated by 60 dB to reduce thermal noise reaching the sample. The response from the readout resonators located on the sample is amplified using a HEMT- and room-temperature amplifiers. Finally, the readout signal is downconverted and digitized at a 100 MHz IF frequency using a Spectrum Instruments m4x PXI card. Using this scheme, qubit spectra are obtained under continuous excitation and readout. Artificial atoms coherence times are characterised using conventional time-domain techniques \cite{schuster2005, krantz2019}.


\begin{figure}
    \centering
    \includegraphics[width=\linewidth]{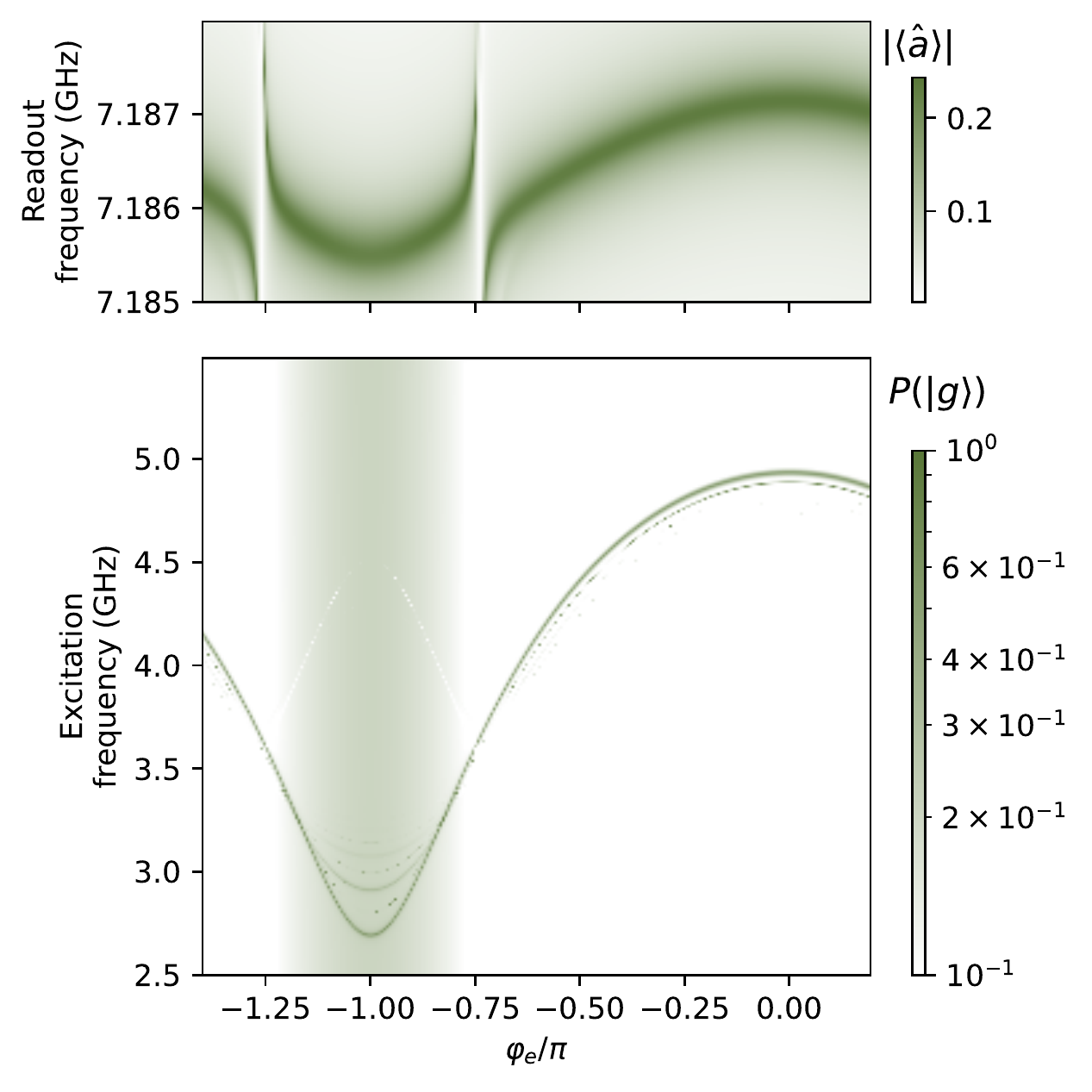}
    \caption{Simulated spectra for the kinemon I, obtained from the master equation solution. Color shows the expectation value of the cavity field $\lvert\left\langle \hat a \right \rangle\lvert$ (top) and the depopulation of the ground state, $1-P(\ket{g})$, (bottom). Simulation parameters: $\Omega/2\pi=200$ MHz for the kinemon spectrum ($=0.1$ MHz for the cavity spectrum), $\omega_r/2\pi= 7.1851$ GHz, $g/2\pi = 64$ MHz, $\kappa=2.5\ \mu$s$^{-1}$, $\gamma =10\, \mu$s$^{-1}$.}
    \label{fig:I_spec_sim}
\end{figure}

\section{Cavity coupling}\label{app:model}

To plot the model curves describing the dependence of the resonator frequency on the magnetic flux (the upper panels of \autoref{fig:spectra}), we solve the following Hamiltonian in the $\varphi$-basis for the kinemon and in the Fock basis for the cavity:
\[
\hat{\mathcal{H}}_\mathrm{cQED} = \hat{\mathcal{H}} + \hbar\omega_r (\hat a^\dag \hat a + 1/2)
+ \hbar g (\hat a^\dag - \hat a) \otimes \frac{\partial }{\partial \varphi},
\]
where $\hat{\mathcal{H}}$ is defined by Eq. \eqref{eq:H_kinemon}, $\hat a$ is the bosonic annihilation operator, and the capacitive coupling term is derived using the canonical quantization expression for the $\hat n$ operator, $[\hat n, \hat \varphi] = i$. The corresponding coupling coefficient $g$ is defined by fitting the model to the data.

For numerical solution of the stationary Shroedinger equation for $\hat{\mathcal{H}}_\text{cQED}$, we use separate finite difference formulas of the 6\textsuperscript{th} order to approximate the first and the second derivatives. This allows to achieve convergence of the necessary low-lying eigenenergies on a rough $\varphi$ grid of 30 to 50 nodes, which then facilitates execution of the computationally-demanding fitting algorithms.

To reproduce the multiphoton transitions which can be observed in the lower panels of \autoref{fig:spectra}, we also perform a time-domain simulation by solving the GKSL equation based on the $\hat{\mathcal{H}}_\mathrm{cQED}$-defined unitary evolution and dissipative energy relaxation dynamics characterized by the collapse operators $\sqrt{\kappa} \hat a$ and $\sqrt{\gamma} \hat b$, $\kappa, \gamma$ being the corresponding decay rates. Here, the kinemon lowering operator $\hat b$ for each value of $\varphi_e$ is constructed from the lowest eigenstates $\ket{E_n}$,
\[
\hat b = \hat{\mathbbm{1}}_\text{r} \otimes \sum_n \ket{E_n}\bra{E_{n+1}}.
\]
Then, at a given $\varphi_e$, the master equation is solved with an addition of a resonator driving term to $\hat{\mathcal{H}}_\mathrm{cQED}$ of the form $\hbar \Omega (\hat a^\dag + \hat a) \cos{\omega_\text{d} t}$. The driving frequency $\omega_\text{d}/2\pi$ is scanned through the same range that is used in the spectroscopy, i.e., from 2.5 to 5.5 GHz, and the resulting steady-state density matrices are saved for further calculation of the observables. 

We show the results for both the transmission spectroscopy and the two-tone spectroscopy in \autoref{fig:I_spec_sim}. We note in overall a good agreement with the experimental data, but find an extra spectral line in \autoref{fig:spectra}(a), having its minimum at around 4.7 GHz, which is not reproduced in the simulation. We attribute it to a sideband two-photon process taking one thermal photon from the cavity and exciting the kinemon to its 4-th excited state located around 12 GHz above the ground state at the lower sweet spot (in the modeling, the cavity it at zero temperature, so such a process can not be observed). The upside-down-looking spectral line having a maximum frequency of 4.5 GHz is a vice-versa process, depopulating the kinemon from $\ket{e}$ to $\ket{g}$ and exciting the cavity. It is observable in the simulation as $\ket{e}$ is slighlty populated for $\varphi \in [-1.2, -0.8]$ due to the numerical errors.

\bibliography{main}

\begin{thebibliography}{47}%
\makeatletter
\providecommand \@ifxundefined [1]{%
 \@ifx{#1\undefined}
}%
\providecommand \@ifnum [1]{%
 \ifnum #1\expandafter \@firstoftwo
 \else \expandafter \@secondoftwo
 \fi
}%
\providecommand \@ifx [1]{%
 \ifx #1\expandafter \@firstoftwo
 \else \expandafter \@secondoftwo
 \fi
}%
\providecommand \natexlab [1]{#1}%
\providecommand \enquote  [1]{``#1''}%
\providecommand \bibnamefont  [1]{#1}%
\providecommand \bibfnamefont [1]{#1}%
\providecommand \citenamefont [1]{#1}%
\providecommand \href@noop [0]{\@secondoftwo}%
\providecommand \href [0]{\begingroup \@sanitize@url \@href}%
\providecommand \@href[1]{\@@startlink{#1}\@@href}%
\providecommand \@@href[1]{\endgroup#1\@@endlink}%
\providecommand \@sanitize@url [0]{\catcode `\\12\catcode `\$12\catcode
  `\&12\catcode `\#12\catcode `\^12\catcode `\_12\catcode `\%12\relax}%
\providecommand \@@startlink[1]{}%
\providecommand \@@endlink[0]{}%
\providecommand \url  [0]{\begingroup\@sanitize@url \@url }%
\providecommand \@url [1]{\endgroup\@href {#1}{\urlprefix }}%
\providecommand \urlprefix  [0]{URL }%
\providecommand \Eprint [0]{\href }%
\providecommand \doibase [0]{https://doi.org/}%
\providecommand \selectlanguage [0]{\@gobble}%
\providecommand \bibinfo  [0]{\@secondoftwo}%
\providecommand \bibfield  [0]{\@secondoftwo}%
\providecommand \translation [1]{[#1]}%
\providecommand \BibitemOpen [0]{}%
\providecommand \bibitemStop [0]{}%
\providecommand \bibitemNoStop [0]{.\EOS\space}%
\providecommand \EOS [0]{\spacefactor3000\relax}%
\providecommand \BibitemShut  [1]{\csname bibitem#1\endcsname}%
\let\auto@bib@innerbib\@empty
\bibitem [{\citenamefont {Koch}\ \emph {et~al.}(2007)\citenamefont {Koch},
  \citenamefont {Yu}, \citenamefont {Gambetta}, \citenamefont {Houck},
  \citenamefont {Schuster}, \citenamefont {Majer}, \citenamefont {Blais},
  \citenamefont {Devoret}, \citenamefont {Girvin},\ and\ \citenamefont
  {Schoelkopf}}]{koch2007}%
  \BibitemOpen
  \bibfield  {author} {\bibinfo {author} {\bibfnamefont {J.}~\bibnamefont
  {Koch}}, \bibinfo {author} {\bibfnamefont {T.~M.}\ \bibnamefont {Yu}},
  \bibinfo {author} {\bibfnamefont {J.}~\bibnamefont {Gambetta}}, \bibinfo
  {author} {\bibfnamefont {A.~A.}\ \bibnamefont {Houck}}, \bibinfo {author}
  {\bibfnamefont {D.~I.}\ \bibnamefont {Schuster}}, \bibinfo {author}
  {\bibfnamefont {J.}~\bibnamefont {Majer}}, \bibinfo {author} {\bibfnamefont
  {A.}~\bibnamefont {Blais}}, \bibinfo {author} {\bibfnamefont {M.~H.}\
  \bibnamefont {Devoret}}, \bibinfo {author} {\bibfnamefont {S.~M.}\
  \bibnamefont {Girvin}},\ and\ \bibinfo {author} {\bibfnamefont {R.~J.}\
  \bibnamefont {Schoelkopf}},\ }\bibfield  {title} {\bibinfo {title}
  {Charge-insensitive qubit design derived from the cooper pair box},\ }\href
  {https://doi.org/10.1103/PhysRevA.76.042319} {\bibfield  {journal} {\bibinfo
  {journal} {Phys. Rev. A}\ }\textbf {\bibinfo {volume} {76}},\ \bibinfo
  {pages} {042319} (\bibinfo {year} {2007})}\BibitemShut {NoStop}%
\bibitem [{\citenamefont {Schreier}\ \emph {et~al.}(2008)\citenamefont
  {Schreier}, \citenamefont {Houck}, \citenamefont {Koch}, \citenamefont
  {Schuster}, \citenamefont {Johnson}, \citenamefont {Chow}, \citenamefont
  {Gambetta}, \citenamefont {Majer}, \citenamefont {Frunzio}, \citenamefont
  {Devoret}, \citenamefont {Girvin},\ and\ \citenamefont
  {Schoelkopf}}]{schreier2008}%
  \BibitemOpen
  \bibfield  {author} {\bibinfo {author} {\bibfnamefont {J.~A.}\ \bibnamefont
  {Schreier}}, \bibinfo {author} {\bibfnamefont {A.~A.}\ \bibnamefont {Houck}},
  \bibinfo {author} {\bibfnamefont {J.}~\bibnamefont {Koch}}, \bibinfo {author}
  {\bibfnamefont {D.~I.}\ \bibnamefont {Schuster}}, \bibinfo {author}
  {\bibfnamefont {B.~R.}\ \bibnamefont {Johnson}}, \bibinfo {author}
  {\bibfnamefont {J.~M.}\ \bibnamefont {Chow}}, \bibinfo {author}
  {\bibfnamefont {J.~M.}\ \bibnamefont {Gambetta}}, \bibinfo {author}
  {\bibfnamefont {J.}~\bibnamefont {Majer}}, \bibinfo {author} {\bibfnamefont
  {L.}~\bibnamefont {Frunzio}}, \bibinfo {author} {\bibfnamefont {M.~H.}\
  \bibnamefont {Devoret}}, \bibinfo {author} {\bibfnamefont {S.~M.}\
  \bibnamefont {Girvin}},\ and\ \bibinfo {author} {\bibfnamefont {R.~J.}\
  \bibnamefont {Schoelkopf}},\ }\bibfield  {title} {\bibinfo {title}
  {Suppressing charge noise decoherence in superconducting charge qubits},\
  }\href {https://doi.org/10.1103/PhysRevB.77.180502} {\bibfield  {journal}
  {\bibinfo  {journal} {Phys. Rev. B}\ }\textbf {\bibinfo {volume} {77}},\
  \bibinfo {pages} {180502} (\bibinfo {year} {2008})}\BibitemShut {NoStop}%
\bibitem [{\citenamefont {Krantz}\ \emph {et~al.}(2019)\citenamefont {Krantz},
  \citenamefont {Kjaergaard}, \citenamefont {Yan}, \citenamefont {Orlando},
  \citenamefont {Gustavsson},\ and\ \citenamefont {Oliver}}]{krantz2019}%
  \BibitemOpen
  \bibfield  {author} {\bibinfo {author} {\bibfnamefont {P.}~\bibnamefont
  {Krantz}}, \bibinfo {author} {\bibfnamefont {M.}~\bibnamefont {Kjaergaard}},
  \bibinfo {author} {\bibfnamefont {F.}~\bibnamefont {Yan}}, \bibinfo {author}
  {\bibfnamefont {T.~P.}\ \bibnamefont {Orlando}}, \bibinfo {author}
  {\bibfnamefont {S.}~\bibnamefont {Gustavsson}},\ and\ \bibinfo {author}
  {\bibfnamefont {W.~D.}\ \bibnamefont {Oliver}},\ }\bibfield  {title}
  {\bibinfo {title} {A quantum engineer's guide to superconducting qubits},\
  }\href {https://doi.org/10.1063/1.5089550} {\bibfield  {journal} {\bibinfo
  {journal} {Applied Physics Reviews}\ }\textbf {\bibinfo {volume} {6}},\
  \bibinfo {pages} {021318} (\bibinfo {year} {2019})}\BibitemShut {NoStop}%
\bibitem [{\citenamefont {Burnett}\ \emph {et~al.}(2019)\citenamefont
  {Burnett}, \citenamefont {Bengtsson}, \citenamefont {Scigliuzzo},
  \citenamefont {Niepce}, \citenamefont {Kudra}, \citenamefont {Delsing},\ and\
  \citenamefont {Bylander}}]{burnett2019}%
  \BibitemOpen
  \bibfield  {author} {\bibinfo {author} {\bibfnamefont {J.~J.}\ \bibnamefont
  {Burnett}}, \bibinfo {author} {\bibfnamefont {A.}~\bibnamefont {Bengtsson}},
  \bibinfo {author} {\bibfnamefont {M.}~\bibnamefont {Scigliuzzo}}, \bibinfo
  {author} {\bibfnamefont {D.}~\bibnamefont {Niepce}}, \bibinfo {author}
  {\bibfnamefont {M.}~\bibnamefont {Kudra}}, \bibinfo {author} {\bibfnamefont
  {P.}~\bibnamefont {Delsing}},\ and\ \bibinfo {author} {\bibfnamefont
  {J.}~\bibnamefont {Bylander}},\ }\bibfield  {title} {\bibinfo {title}
  {Decoherence benchmarking of superconducting qubits},\ }\href
  {https://doi.org/10.1038/s41534-019-0168-5} {\bibfield  {journal} {\bibinfo
  {journal} {npj Quantum Information}\ }\textbf {\bibinfo {volume} {5}},\
  \bibinfo {pages} {54} (\bibinfo {year} {2019})}\BibitemShut {NoStop}%
\bibitem [{\citenamefont {Schutjens}\ \emph {et~al.}(2013)\citenamefont
  {Schutjens}, \citenamefont {Dagga}, \citenamefont {Egger},\ and\
  \citenamefont {Wilhelm}}]{schutjens2013}%
  \BibitemOpen
  \bibfield  {author} {\bibinfo {author} {\bibfnamefont {R.}~\bibnamefont
  {Schutjens}}, \bibinfo {author} {\bibfnamefont {F.~A.}\ \bibnamefont
  {Dagga}}, \bibinfo {author} {\bibfnamefont {D.~J.}\ \bibnamefont {Egger}},\
  and\ \bibinfo {author} {\bibfnamefont {F.~K.}\ \bibnamefont {Wilhelm}},\
  }\bibfield  {title} {\bibinfo {title} {Single-qubit gates in
  frequency-crowded transmon systems},\ }\href
  {https://doi.org/10.1103/PhysRevA.88.052330} {\bibfield  {journal} {\bibinfo
  {journal} {Phys. Rev. A}\ }\textbf {\bibinfo {volume} {88}},\ \bibinfo
  {pages} {052330} (\bibinfo {year} {2013})}\BibitemShut {NoStop}%
\bibitem [{\citenamefont {Vesterinen}\ \emph {et~al.}(2014)\citenamefont
  {Vesterinen}, \citenamefont {Saira}, \citenamefont {Bruno},\ and\
  \citenamefont {DiCarlo}}]{vesterinen2014}%
  \BibitemOpen
  \bibfield  {author} {\bibinfo {author} {\bibfnamefont {V.}~\bibnamefont
  {Vesterinen}}, \bibinfo {author} {\bibfnamefont {O.~P.}\ \bibnamefont
  {Saira}}, \bibinfo {author} {\bibfnamefont {A.}~\bibnamefont {Bruno}},\ and\
  \bibinfo {author} {\bibfnamefont {L.}~\bibnamefont {DiCarlo}},\ }\href
  {https://doi.org/10.48550/ARXIV.1405.0450} {\bibinfo {title} {Mitigating
  information leakage in a crowded spectrum of weakly anharmonic qubits}}
  (\bibinfo {year} {2014})\BibitemShut {NoStop}%
\bibitem [{\citenamefont {Chen}\ \emph {et~al.}(2016)\citenamefont {Chen},
  \citenamefont {Kelly}, \citenamefont {Quintana}, \citenamefont {Barends},
  \citenamefont {Campbell}, \citenamefont {Chen}, \citenamefont {Chiaro},
  \citenamefont {Dunsworth}, \citenamefont {Fowler}, \citenamefont {Lucero},
  \citenamefont {Jeffrey}, \citenamefont {Megrant}, \citenamefont {Mutus},
  \citenamefont {Neeley}, \citenamefont {Neill}, \citenamefont {O'Malley},
  \citenamefont {Roushan}, \citenamefont {Sank}, \citenamefont {Vainsencher},
  \citenamefont {Wenner}, \citenamefont {White}, \citenamefont {Korotkov},\
  and\ \citenamefont {Martinis}}]{chen2016}%
  \BibitemOpen
  \bibfield  {author} {\bibinfo {author} {\bibfnamefont {Z.}~\bibnamefont
  {Chen}}, \bibinfo {author} {\bibfnamefont {J.}~\bibnamefont {Kelly}},
  \bibinfo {author} {\bibfnamefont {C.}~\bibnamefont {Quintana}}, \bibinfo
  {author} {\bibfnamefont {R.}~\bibnamefont {Barends}}, \bibinfo {author}
  {\bibfnamefont {B.}~\bibnamefont {Campbell}}, \bibinfo {author}
  {\bibfnamefont {Y.}~\bibnamefont {Chen}}, \bibinfo {author} {\bibfnamefont
  {B.}~\bibnamefont {Chiaro}}, \bibinfo {author} {\bibfnamefont
  {A.}~\bibnamefont {Dunsworth}}, \bibinfo {author} {\bibfnamefont {A.~G.}\
  \bibnamefont {Fowler}}, \bibinfo {author} {\bibfnamefont {E.}~\bibnamefont
  {Lucero}}, \bibinfo {author} {\bibfnamefont {E.}~\bibnamefont {Jeffrey}},
  \bibinfo {author} {\bibfnamefont {A.}~\bibnamefont {Megrant}}, \bibinfo
  {author} {\bibfnamefont {J.}~\bibnamefont {Mutus}}, \bibinfo {author}
  {\bibfnamefont {M.}~\bibnamefont {Neeley}}, \bibinfo {author} {\bibfnamefont
  {C.}~\bibnamefont {Neill}}, \bibinfo {author} {\bibfnamefont {P.~J.~J.}\
  \bibnamefont {O'Malley}}, \bibinfo {author} {\bibfnamefont {P.}~\bibnamefont
  {Roushan}}, \bibinfo {author} {\bibfnamefont {D.}~\bibnamefont {Sank}},
  \bibinfo {author} {\bibfnamefont {A.}~\bibnamefont {Vainsencher}}, \bibinfo
  {author} {\bibfnamefont {J.}~\bibnamefont {Wenner}}, \bibinfo {author}
  {\bibfnamefont {T.~C.}\ \bibnamefont {White}}, \bibinfo {author}
  {\bibfnamefont {A.~N.}\ \bibnamefont {Korotkov}},\ and\ \bibinfo {author}
  {\bibfnamefont {J.~M.}\ \bibnamefont {Martinis}},\ }\bibfield  {title}
  {\bibinfo {title} {Measuring and suppressing quantum state leakage in a
  superconducting qubit},\ }\href
  {https://doi.org/10.1103/PhysRevLett.116.020501} {\bibfield  {journal}
  {\bibinfo  {journal} {Phys. Rev. Lett.}\ }\textbf {\bibinfo {volume} {116}},\
  \bibinfo {pages} {020501} (\bibinfo {year} {2016})}\BibitemShut {NoStop}%
\bibitem [{\citenamefont {McEwen}\ \emph {et~al.}(2021)\citenamefont {McEwen},
  \citenamefont {Kafri}, \citenamefont {Chen}, \citenamefont {Atalaya},
  \citenamefont {Satzinger}, \citenamefont {Quintana}, \citenamefont {Klimov},
  \citenamefont {Sank}, \citenamefont {Gidney}, \citenamefont {Fowler},
  \citenamefont {Arute}, \citenamefont {Arya}, \citenamefont {Buckley},
  \citenamefont {Burkett}, \citenamefont {Bushnell}, \citenamefont {Chiaro},
  \citenamefont {Collins}, \citenamefont {Demura}, \citenamefont {Dunsworth},
  \citenamefont {Erickson}, \citenamefont {Foxen}, \citenamefont {Giustina},
  \citenamefont {Huang}, \citenamefont {Hong}, \citenamefont {Jeffrey},
  \citenamefont {Kim}, \citenamefont {Kechedzhi}, \citenamefont {Kostritsa},
  \citenamefont {Laptev}, \citenamefont {Megrant}, \citenamefont {Mi},
  \citenamefont {Mutus}, \citenamefont {Naaman}, \citenamefont {Neeley},
  \citenamefont {Neill}, \citenamefont {Niu}, \citenamefont {Paler},
  \citenamefont {Redd}, \citenamefont {Roushan}, \citenamefont {White},
  \citenamefont {Yao}, \citenamefont {Yeh}, \citenamefont {Zalcman},
  \citenamefont {Chen}, \citenamefont {Smelyanskiy}, \citenamefont {Martinis},
  \citenamefont {Neven}, \citenamefont {Kelly}, \citenamefont {Korotkov},
  \citenamefont {Petukhov},\ and\ \citenamefont {Barends}}]{mcewen2021}%
  \BibitemOpen
  \bibfield  {author} {\bibinfo {author} {\bibfnamefont {M.}~\bibnamefont
  {McEwen}}, \bibinfo {author} {\bibfnamefont {D.}~\bibnamefont {Kafri}},
  \bibinfo {author} {\bibfnamefont {Z.}~\bibnamefont {Chen}}, \bibinfo {author}
  {\bibfnamefont {J.}~\bibnamefont {Atalaya}}, \bibinfo {author} {\bibfnamefont
  {K.~J.}\ \bibnamefont {Satzinger}}, \bibinfo {author} {\bibfnamefont
  {C.}~\bibnamefont {Quintana}}, \bibinfo {author} {\bibfnamefont {P.~V.}\
  \bibnamefont {Klimov}}, \bibinfo {author} {\bibfnamefont {D.}~\bibnamefont
  {Sank}}, \bibinfo {author} {\bibfnamefont {C.}~\bibnamefont {Gidney}},
  \bibinfo {author} {\bibfnamefont {A.~G.}\ \bibnamefont {Fowler}}, \bibinfo
  {author} {\bibfnamefont {F.}~\bibnamefont {Arute}}, \bibinfo {author}
  {\bibfnamefont {K.}~\bibnamefont {Arya}}, \bibinfo {author} {\bibfnamefont
  {B.}~\bibnamefont {Buckley}}, \bibinfo {author} {\bibfnamefont
  {B.}~\bibnamefont {Burkett}}, \bibinfo {author} {\bibfnamefont
  {N.}~\bibnamefont {Bushnell}}, \bibinfo {author} {\bibfnamefont
  {B.}~\bibnamefont {Chiaro}}, \bibinfo {author} {\bibfnamefont
  {R.}~\bibnamefont {Collins}}, \bibinfo {author} {\bibfnamefont
  {S.}~\bibnamefont {Demura}}, \bibinfo {author} {\bibfnamefont
  {A.}~\bibnamefont {Dunsworth}}, \bibinfo {author} {\bibfnamefont
  {C.}~\bibnamefont {Erickson}}, \bibinfo {author} {\bibfnamefont
  {B.}~\bibnamefont {Foxen}}, \bibinfo {author} {\bibfnamefont
  {M.}~\bibnamefont {Giustina}}, \bibinfo {author} {\bibfnamefont
  {T.}~\bibnamefont {Huang}}, \bibinfo {author} {\bibfnamefont
  {S.}~\bibnamefont {Hong}}, \bibinfo {author} {\bibfnamefont {E.}~\bibnamefont
  {Jeffrey}}, \bibinfo {author} {\bibfnamefont {S.}~\bibnamefont {Kim}},
  \bibinfo {author} {\bibfnamefont {K.}~\bibnamefont {Kechedzhi}}, \bibinfo
  {author} {\bibfnamefont {F.}~\bibnamefont {Kostritsa}}, \bibinfo {author}
  {\bibfnamefont {P.}~\bibnamefont {Laptev}}, \bibinfo {author} {\bibfnamefont
  {A.}~\bibnamefont {Megrant}}, \bibinfo {author} {\bibfnamefont
  {X.}~\bibnamefont {Mi}}, \bibinfo {author} {\bibfnamefont {J.}~\bibnamefont
  {Mutus}}, \bibinfo {author} {\bibfnamefont {O.}~\bibnamefont {Naaman}},
  \bibinfo {author} {\bibfnamefont {M.}~\bibnamefont {Neeley}}, \bibinfo
  {author} {\bibfnamefont {C.}~\bibnamefont {Neill}}, \bibinfo {author}
  {\bibfnamefont {M.}~\bibnamefont {Niu}}, \bibinfo {author} {\bibfnamefont
  {A.}~\bibnamefont {Paler}}, \bibinfo {author} {\bibfnamefont
  {N.}~\bibnamefont {Redd}}, \bibinfo {author} {\bibfnamefont {P.}~\bibnamefont
  {Roushan}}, \bibinfo {author} {\bibfnamefont {T.~C.}\ \bibnamefont {White}},
  \bibinfo {author} {\bibfnamefont {J.}~\bibnamefont {Yao}}, \bibinfo {author}
  {\bibfnamefont {P.}~\bibnamefont {Yeh}}, \bibinfo {author} {\bibfnamefont
  {A.}~\bibnamefont {Zalcman}}, \bibinfo {author} {\bibfnamefont
  {Y.}~\bibnamefont {Chen}}, \bibinfo {author} {\bibfnamefont {V.~N.}\
  \bibnamefont {Smelyanskiy}}, \bibinfo {author} {\bibfnamefont {J.~M.}\
  \bibnamefont {Martinis}}, \bibinfo {author} {\bibfnamefont {H.}~\bibnamefont
  {Neven}}, \bibinfo {author} {\bibfnamefont {J.}~\bibnamefont {Kelly}},
  \bibinfo {author} {\bibfnamefont {A.~N.}\ \bibnamefont {Korotkov}}, \bibinfo
  {author} {\bibfnamefont {A.~G.}\ \bibnamefont {Petukhov}},\ and\ \bibinfo
  {author} {\bibfnamefont {R.}~\bibnamefont {Barends}},\ }\bibfield  {title}
  {\bibinfo {title} {Removing leakage-induced correlated errors in
  superconducting quantum error correction},\ }\href
  {https://doi.org/10.1038/s41467-021-21982-y} {\bibfield  {journal} {\bibinfo
  {journal} {Nature Communications}\ }\textbf {\bibinfo {volume} {12}},\
  \bibinfo {pages} {1761} (\bibinfo {year} {2021})}\BibitemShut {NoStop}%
\bibitem [{\citenamefont {Bultink}\ \emph {et~al.}(2020)\citenamefont
  {Bultink}, \citenamefont {O'Brien}, \citenamefont {Vollmer}, \citenamefont
  {Muthusubramanian}, \citenamefont {Beekman}, \citenamefont {Rol},
  \citenamefont {Fu}, \citenamefont {Tarasinski}, \citenamefont {Ostroukh},
  \citenamefont {Varbanov}, \citenamefont {Bruno},\ and\ \citenamefont
  {DiCarlo}}]{bultink2020}%
  \BibitemOpen
  \bibfield  {author} {\bibinfo {author} {\bibfnamefont {C.~C.}\ \bibnamefont
  {Bultink}}, \bibinfo {author} {\bibfnamefont {T.~E.}\ \bibnamefont
  {O'Brien}}, \bibinfo {author} {\bibfnamefont {R.}~\bibnamefont {Vollmer}},
  \bibinfo {author} {\bibfnamefont {N.}~\bibnamefont {Muthusubramanian}},
  \bibinfo {author} {\bibfnamefont {M.~W.}\ \bibnamefont {Beekman}}, \bibinfo
  {author} {\bibfnamefont {M.~A.}\ \bibnamefont {Rol}}, \bibinfo {author}
  {\bibfnamefont {X.}~\bibnamefont {Fu}}, \bibinfo {author} {\bibfnamefont
  {B.}~\bibnamefont {Tarasinski}}, \bibinfo {author} {\bibfnamefont
  {V.}~\bibnamefont {Ostroukh}}, \bibinfo {author} {\bibfnamefont
  {B.}~\bibnamefont {Varbanov}}, \bibinfo {author} {\bibfnamefont
  {A.}~\bibnamefont {Bruno}},\ and\ \bibinfo {author} {\bibfnamefont
  {L.}~\bibnamefont {DiCarlo}},\ }\bibfield  {title} {\bibinfo {title}
  {Protecting quantum entanglement from leakage and qubit errors via repetitive
  parity measurements},\ }\href {https://doi.org/10.1126/sciadv.aay3050}
  {\bibfield  {journal} {\bibinfo  {journal} {Science Advances}\ }\textbf
  {\bibinfo {volume} {6}},\ \bibinfo {pages} {eaay3050} (\bibinfo {year}
  {2020})},\ \Eprint
  {https://arxiv.org/abs/https://www.science.org/doi/pdf/10.1126/sciadv.aay3050}
  {https://www.science.org/doi/pdf/10.1126/sciadv.aay3050} \BibitemShut
  {NoStop}%
\bibitem [{\citenamefont {Miao}\ \emph {et~al.}(2022)\citenamefont {Miao} \emph
  {et~al.}}]{miao2022}%
  \BibitemOpen
  \bibfield  {author} {\bibinfo {author} {\bibfnamefont {K.~C.}\ \bibnamefont
  {Miao}} \emph {et~al.},\ }\href {https://doi.org/10.48550/arXiv.2211.04728}
  {\bibinfo {title} {Overcoming leakage in scalable quantum error correction}}
  (\bibinfo {year} {2022}),\ \Eprint {https://arxiv.org/abs/2211.04728}
  {arXiv:2211.04728 [quant-ph]} \BibitemShut {NoStop}%
\bibitem [{\citenamefont {Roy}\ \emph {et~al.}(2022)\citenamefont {Roy},
  \citenamefont {Li}, \citenamefont {Kapit},\ and\ \citenamefont
  {Schuster}}]{roy2022}%
  \BibitemOpen
  \bibfield  {author} {\bibinfo {author} {\bibfnamefont {T.}~\bibnamefont
  {Roy}}, \bibinfo {author} {\bibfnamefont {Z.}~\bibnamefont {Li}}, \bibinfo
  {author} {\bibfnamefont {E.}~\bibnamefont {Kapit}},\ and\ \bibinfo {author}
  {\bibfnamefont {D.~I.}\ \bibnamefont {Schuster}},\ }\href@noop {} {\bibinfo
  {title} {Realization of two-qutrit quantum algorithms on a programmable
  superconducting processor}} (\bibinfo {year} {2022}),\ \Eprint
  {https://arxiv.org/abs/2211.06523} {arXiv:2211.06523 [quant-ph]} \BibitemShut
  {NoStop}%
\bibitem [{\citenamefont {Pechenezhskiy}\ \emph {et~al.}(2020)\citenamefont
  {Pechenezhskiy}, \citenamefont {Mencia}, \citenamefont {Nguyen},
  \citenamefont {Lin},\ and\ \citenamefont {Manucharyan}}]{pechenezhskiy2020}%
  \BibitemOpen
  \bibfield  {author} {\bibinfo {author} {\bibfnamefont {I.~V.}\ \bibnamefont
  {Pechenezhskiy}}, \bibinfo {author} {\bibfnamefont {R.~A.}\ \bibnamefont
  {Mencia}}, \bibinfo {author} {\bibfnamefont {L.~B.}\ \bibnamefont {Nguyen}},
  \bibinfo {author} {\bibfnamefont {Y.-H.}\ \bibnamefont {Lin}},\ and\ \bibinfo
  {author} {\bibfnamefont {V.~E.}\ \bibnamefont {Manucharyan}},\ }\bibfield
  {title} {\bibinfo {title} {The superconducting quasicharge qubit},\ }\href
  {https://doi.org/10.1038/s41586-020-2687-9} {\bibfield  {journal} {\bibinfo
  {journal} {Nature}\ }\textbf {\bibinfo {volume} {585}},\ \bibinfo {pages}
  {368} (\bibinfo {year} {2020})}\BibitemShut {NoStop}%
\bibitem [{\citenamefont {Gyenis}\ \emph {et~al.}(2021)\citenamefont {Gyenis},
  \citenamefont {Mundada}, \citenamefont {Di~Paolo}, \citenamefont {Hazard},
  \citenamefont {You}, \citenamefont {Schuster}, \citenamefont {Koch},
  \citenamefont {Blais},\ and\ \citenamefont {Houck}}]{gyenis2021}%
  \BibitemOpen
  \bibfield  {author} {\bibinfo {author} {\bibfnamefont {A.}~\bibnamefont
  {Gyenis}}, \bibinfo {author} {\bibfnamefont {P.~S.}\ \bibnamefont {Mundada}},
  \bibinfo {author} {\bibfnamefont {A.}~\bibnamefont {Di~Paolo}}, \bibinfo
  {author} {\bibfnamefont {T.~M.}\ \bibnamefont {Hazard}}, \bibinfo {author}
  {\bibfnamefont {X.}~\bibnamefont {You}}, \bibinfo {author} {\bibfnamefont
  {D.~I.}\ \bibnamefont {Schuster}}, \bibinfo {author} {\bibfnamefont
  {J.}~\bibnamefont {Koch}}, \bibinfo {author} {\bibfnamefont {A.}~\bibnamefont
  {Blais}},\ and\ \bibinfo {author} {\bibfnamefont {A.~A.}\ \bibnamefont
  {Houck}},\ }\bibfield  {title} {\bibinfo {title} {Experimental realization of
  a protected superconducting circuit derived from the $0$--$\ensuremath{\pi}$
  qubit},\ }\href {https://doi.org/10.1103/PRXQuantum.2.010339} {\bibfield
  {journal} {\bibinfo  {journal} {PRX Quantum}\ }\textbf {\bibinfo {volume}
  {2}},\ \bibinfo {pages} {010339} (\bibinfo {year} {2021})}\BibitemShut
  {NoStop}%
\bibitem [{\citenamefont {Hyypp{\"a}}\ \emph {et~al.}(2022)\citenamefont
  {Hyypp{\"a}}, \citenamefont {Kundu}, \citenamefont {Chan}, \citenamefont
  {Gunyh{\'o}}, \citenamefont {Hotari}, \citenamefont {Janzso}, \citenamefont
  {Juliusson}, \citenamefont {Kiuru}, \citenamefont {Kotilahti}, \citenamefont
  {Landra}, \citenamefont {Liu}, \citenamefont {Marxer}, \citenamefont
  {M{\"a}kinen}, \citenamefont {Orgiazzi}, \citenamefont {Palma}, \citenamefont
  {Savytskyi}, \citenamefont {Tosto}, \citenamefont {Tuorila}, \citenamefont
  {Vadimov}, \citenamefont {Li}, \citenamefont {Ockeloen-Korppi}, \citenamefont
  {Heinsoo}, \citenamefont {Tan}, \citenamefont {Hassel},\ and\ \citenamefont
  {M{\"o}tt{\"o}nen}}]{hyyppa2022}%
  \BibitemOpen
  \bibfield  {author} {\bibinfo {author} {\bibfnamefont {E.}~\bibnamefont
  {Hyypp{\"a}}}, \bibinfo {author} {\bibfnamefont {S.}~\bibnamefont {Kundu}},
  \bibinfo {author} {\bibfnamefont {C.~F.}\ \bibnamefont {Chan}}, \bibinfo
  {author} {\bibfnamefont {A.}~\bibnamefont {Gunyh{\'o}}}, \bibinfo {author}
  {\bibfnamefont {J.}~\bibnamefont {Hotari}}, \bibinfo {author} {\bibfnamefont
  {D.}~\bibnamefont {Janzso}}, \bibinfo {author} {\bibfnamefont
  {K.}~\bibnamefont {Juliusson}}, \bibinfo {author} {\bibfnamefont
  {O.}~\bibnamefont {Kiuru}}, \bibinfo {author} {\bibfnamefont
  {J.}~\bibnamefont {Kotilahti}}, \bibinfo {author} {\bibfnamefont
  {A.}~\bibnamefont {Landra}}, \bibinfo {author} {\bibfnamefont
  {W.}~\bibnamefont {Liu}}, \bibinfo {author} {\bibfnamefont {F.}~\bibnamefont
  {Marxer}}, \bibinfo {author} {\bibfnamefont {A.}~\bibnamefont {M{\"a}kinen}},
  \bibinfo {author} {\bibfnamefont {J.-L.}\ \bibnamefont {Orgiazzi}}, \bibinfo
  {author} {\bibfnamefont {M.}~\bibnamefont {Palma}}, \bibinfo {author}
  {\bibfnamefont {M.}~\bibnamefont {Savytskyi}}, \bibinfo {author}
  {\bibfnamefont {F.}~\bibnamefont {Tosto}}, \bibinfo {author} {\bibfnamefont
  {J.}~\bibnamefont {Tuorila}}, \bibinfo {author} {\bibfnamefont
  {V.}~\bibnamefont {Vadimov}}, \bibinfo {author} {\bibfnamefont
  {T.}~\bibnamefont {Li}}, \bibinfo {author} {\bibfnamefont {C.}~\bibnamefont
  {Ockeloen-Korppi}}, \bibinfo {author} {\bibfnamefont {J.}~\bibnamefont
  {Heinsoo}}, \bibinfo {author} {\bibfnamefont {K.~Y.}\ \bibnamefont {Tan}},
  \bibinfo {author} {\bibfnamefont {J.}~\bibnamefont {Hassel}},\ and\ \bibinfo
  {author} {\bibfnamefont {M.}~\bibnamefont {M{\"o}tt{\"o}nen}},\ }\bibfield
  {title} {\bibinfo {title} {Unimon qubit},\ }\href
  {https://doi.org/10.1038/s41467-022-34614-w} {\bibfield  {journal} {\bibinfo
  {journal} {Nature Communications}\ }\textbf {\bibinfo {volume} {13}},\
  \bibinfo {pages} {6895} (\bibinfo {year} {2022})}\BibitemShut {NoStop}%
\bibitem [{\citenamefont {Hassani}\ \emph {et~al.}(2022)\citenamefont
  {Hassani}, \citenamefont {Peruzzo}, \citenamefont {Kapoor}, \citenamefont
  {Trioni}, \citenamefont {Zemlicka},\ and\ \citenamefont
  {Fink}}]{hassani2022}%
  \BibitemOpen
  \bibfield  {author} {\bibinfo {author} {\bibfnamefont {F.}~\bibnamefont
  {Hassani}}, \bibinfo {author} {\bibfnamefont {M.}~\bibnamefont {Peruzzo}},
  \bibinfo {author} {\bibfnamefont {L.~N.}\ \bibnamefont {Kapoor}}, \bibinfo
  {author} {\bibfnamefont {A.}~\bibnamefont {Trioni}}, \bibinfo {author}
  {\bibfnamefont {M.}~\bibnamefont {Zemlicka}},\ and\ \bibinfo {author}
  {\bibfnamefont {J.~M.}\ \bibnamefont {Fink}},\ }\href@noop {} {\bibinfo
  {title} {A superconducting qubit with noise-insensitive plasmon levels and
  decay-protected fluxon states}} (\bibinfo {year} {2022}),\ \Eprint
  {https://arxiv.org/abs/2202.13917} {arXiv:2202.13917 [cond-mat.mes-hall]}
  \BibitemShut {NoStop}%
\bibitem [{\citenamefont {Orlando}\ \emph {et~al.}(1999)\citenamefont
  {Orlando}, \citenamefont {Mooij}, \citenamefont {Tian}, \citenamefont
  {van~der Wal}, \citenamefont {Levitov}, \citenamefont {Lloyd},\ and\
  \citenamefont {Mazo}}]{orlando1999}%
  \BibitemOpen
  \bibfield  {author} {\bibinfo {author} {\bibfnamefont {T.~P.}\ \bibnamefont
  {Orlando}}, \bibinfo {author} {\bibfnamefont {J.~E.}\ \bibnamefont {Mooij}},
  \bibinfo {author} {\bibfnamefont {L.}~\bibnamefont {Tian}}, \bibinfo {author}
  {\bibfnamefont {C.~H.}\ \bibnamefont {van~der Wal}}, \bibinfo {author}
  {\bibfnamefont {L.~S.}\ \bibnamefont {Levitov}}, \bibinfo {author}
  {\bibfnamefont {S.}~\bibnamefont {Lloyd}},\ and\ \bibinfo {author}
  {\bibfnamefont {J.~J.}\ \bibnamefont {Mazo}},\ }\bibfield  {title} {\bibinfo
  {title} {Superconducting persistent-current qubit},\ }\href
  {https://doi.org/10.1103/PhysRevB.60.15398} {\bibfield  {journal} {\bibinfo
  {journal} {Phys. Rev. B}\ }\textbf {\bibinfo {volume} {60}},\ \bibinfo
  {pages} {15398} (\bibinfo {year} {1999})}\BibitemShut {NoStop}%
\bibitem [{\citenamefont {Chiorescu}\ \emph {et~al.}(2003)\citenamefont
  {Chiorescu}, \citenamefont {Nakamura}, \citenamefont {Harmans},\ and\
  \citenamefont {Mooij}}]{chiorescu2003}%
  \BibitemOpen
  \bibfield  {author} {\bibinfo {author} {\bibfnamefont {I.}~\bibnamefont
  {Chiorescu}}, \bibinfo {author} {\bibfnamefont {Y.}~\bibnamefont {Nakamura}},
  \bibinfo {author} {\bibfnamefont {C.~J. P.~M.}\ \bibnamefont {Harmans}},\
  and\ \bibinfo {author} {\bibfnamefont {J.~E.}\ \bibnamefont {Mooij}},\
  }\bibfield  {title} {\bibinfo {title} {Coherent quantum dynamics of a
  superconducting flux qubit},\ }\href
  {https://doi.org/10.1126/science.1081045} {\bibfield  {journal} {\bibinfo
  {journal} {Science}\ }\textbf {\bibinfo {volume} {299}},\ \bibinfo {pages}
  {1869} (\bibinfo {year} {2003})}\BibitemShut {NoStop}%
\bibitem [{\citenamefont {You}\ \emph {et~al.}(2007)\citenamefont {You},
  \citenamefont {Hu}, \citenamefont {Ashhab},\ and\ \citenamefont
  {Nori}}]{you2007}%
  \BibitemOpen
  \bibfield  {author} {\bibinfo {author} {\bibfnamefont {J.~Q.}\ \bibnamefont
  {You}}, \bibinfo {author} {\bibfnamefont {X.}~\bibnamefont {Hu}}, \bibinfo
  {author} {\bibfnamefont {S.}~\bibnamefont {Ashhab}},\ and\ \bibinfo {author}
  {\bibfnamefont {F.}~\bibnamefont {Nori}},\ }\bibfield  {title} {\bibinfo
  {title} {Low-decoherence flux qubit},\ }\href
  {https://doi.org/10.1103/PhysRevB.75.140515} {\bibfield  {journal} {\bibinfo
  {journal} {Phys. Rev. B}\ }\textbf {\bibinfo {volume} {75}},\ \bibinfo
  {pages} {140515} (\bibinfo {year} {2007})}\BibitemShut {NoStop}%
\bibitem [{\citenamefont {Yan}\ \emph {et~al.}(2016)\citenamefont {Yan},
  \citenamefont {Gustavsson}, \citenamefont {Kamal}, \citenamefont {Birenbaum},
  \citenamefont {Sears}, \citenamefont {Hover}, \citenamefont {Gudmundsen},
  \citenamefont {Rosenberg}, \citenamefont {Samach}, \citenamefont {Weber},
  \citenamefont {Yoder}, \citenamefont {Orlando}, \citenamefont {Clarke},
  \citenamefont {Kerman},\ and\ \citenamefont {Oliver}}]{yan2016}%
  \BibitemOpen
  \bibfield  {author} {\bibinfo {author} {\bibfnamefont {F.}~\bibnamefont
  {Yan}}, \bibinfo {author} {\bibfnamefont {S.}~\bibnamefont {Gustavsson}},
  \bibinfo {author} {\bibfnamefont {A.}~\bibnamefont {Kamal}}, \bibinfo
  {author} {\bibfnamefont {J.}~\bibnamefont {Birenbaum}}, \bibinfo {author}
  {\bibfnamefont {A.~P.}\ \bibnamefont {Sears}}, \bibinfo {author}
  {\bibfnamefont {D.}~\bibnamefont {Hover}}, \bibinfo {author} {\bibfnamefont
  {T.~J.}\ \bibnamefont {Gudmundsen}}, \bibinfo {author} {\bibfnamefont
  {D.}~\bibnamefont {Rosenberg}}, \bibinfo {author} {\bibfnamefont
  {G.}~\bibnamefont {Samach}}, \bibinfo {author} {\bibfnamefont
  {S.}~\bibnamefont {Weber}}, \bibinfo {author} {\bibfnamefont {J.~L.}\
  \bibnamefont {Yoder}}, \bibinfo {author} {\bibfnamefont {T.~P.}\ \bibnamefont
  {Orlando}}, \bibinfo {author} {\bibfnamefont {J.}~\bibnamefont {Clarke}},
  \bibinfo {author} {\bibfnamefont {A.~J.}\ \bibnamefont {Kerman}},\ and\
  \bibinfo {author} {\bibfnamefont {W.~D.}\ \bibnamefont {Oliver}},\ }\bibfield
   {title} {\bibinfo {title} {The flux qubit revisited to enhance coherence and
  reproducibility},\ }\href {https://doi.org/10.1038/ncomms12964} {\bibfield
  {journal} {\bibinfo  {journal} {Nature Communications}\ }\textbf {\bibinfo
  {volume} {7}},\ \bibinfo {pages} {12964} (\bibinfo {year}
  {2016})}\BibitemShut {NoStop}%
\bibitem [{\citenamefont {Peltonen}\ \emph {et~al.}(2018)\citenamefont
  {Peltonen}, \citenamefont {Coumou}, \citenamefont {Peng}, \citenamefont
  {Klapwijk}, \citenamefont {Tsai},\ and\ \citenamefont
  {Astafiev}}]{peltonen2018}%
  \BibitemOpen
  \bibfield  {author} {\bibinfo {author} {\bibfnamefont {J.~T.}\ \bibnamefont
  {Peltonen}}, \bibinfo {author} {\bibfnamefont {P.~C. J.~J.}\ \bibnamefont
  {Coumou}}, \bibinfo {author} {\bibfnamefont {Z.~H.}\ \bibnamefont {Peng}},
  \bibinfo {author} {\bibfnamefont {T.~M.}\ \bibnamefont {Klapwijk}}, \bibinfo
  {author} {\bibfnamefont {J.~S.}\ \bibnamefont {Tsai}},\ and\ \bibinfo
  {author} {\bibfnamefont {O.~V.}\ \bibnamefont {Astafiev}},\ }\bibfield
  {title} {\bibinfo {title} {Hybrid rf squid qubit based on high kinetic
  inductance},\ }\href {https://doi.org/10.1038/s41598-018-27154-1} {\bibfield
  {journal} {\bibinfo  {journal} {Scientific Reports}\ }\textbf {\bibinfo
  {volume} {8}},\ \bibinfo {pages} {10033} (\bibinfo {year}
  {2018})}\BibitemShut {NoStop}%
\bibitem [{\citenamefont {Moskalenko}\ \emph {et~al.}(2022)\citenamefont
  {Moskalenko}, \citenamefont {Simakov}, \citenamefont {Abramov}, \citenamefont
  {Grigorev}, \citenamefont {Moskalev}, \citenamefont {Pishchimova},
  \citenamefont {Smirnov}, \citenamefont {Zikiy}, \citenamefont {Rodionov},\
  and\ \citenamefont {Besedin}}]{moskalenko2022}%
  \BibitemOpen
  \bibfield  {author} {\bibinfo {author} {\bibfnamefont {I.~N.}\ \bibnamefont
  {Moskalenko}}, \bibinfo {author} {\bibfnamefont {I.~A.}\ \bibnamefont
  {Simakov}}, \bibinfo {author} {\bibfnamefont {N.~N.}\ \bibnamefont
  {Abramov}}, \bibinfo {author} {\bibfnamefont {A.~A.}\ \bibnamefont
  {Grigorev}}, \bibinfo {author} {\bibfnamefont {D.~O.}\ \bibnamefont
  {Moskalev}}, \bibinfo {author} {\bibfnamefont {A.~A.}\ \bibnamefont
  {Pishchimova}}, \bibinfo {author} {\bibfnamefont {N.~S.}\ \bibnamefont
  {Smirnov}}, \bibinfo {author} {\bibfnamefont {E.~V.}\ \bibnamefont {Zikiy}},
  \bibinfo {author} {\bibfnamefont {I.~A.}\ \bibnamefont {Rodionov}},\ and\
  \bibinfo {author} {\bibfnamefont {I.~S.}\ \bibnamefont {Besedin}},\
  }\bibfield  {title} {\bibinfo {title} {High fidelity two-qubit gates on
  fluxoniums using a tunable coupler},\ }\bibfield  {journal} {\bibinfo
  {journal} {npj Quantum Information}\ }\textbf {\bibinfo {volume} {8}},\ \href
  {https://doi.org/10.1038/s41534-022-00644-x} {10.1038/s41534-022-00644-x}
  (\bibinfo {year} {2022})\BibitemShut {NoStop}%
\bibitem [{\citenamefont {Paauw}\ \emph {et~al.}(2009)\citenamefont {Paauw},
  \citenamefont {Fedorov}, \citenamefont {Harmans},\ and\ \citenamefont
  {Mooij}}]{paauw2009}%
  \BibitemOpen
  \bibfield  {author} {\bibinfo {author} {\bibfnamefont {F.~G.}\ \bibnamefont
  {Paauw}}, \bibinfo {author} {\bibfnamefont {A.}~\bibnamefont {Fedorov}},
  \bibinfo {author} {\bibfnamefont {C.~J. P.~M.}\ \bibnamefont {Harmans}},\
  and\ \bibinfo {author} {\bibfnamefont {J.~E.}\ \bibnamefont {Mooij}},\
  }\bibfield  {title} {\bibinfo {title} {Tuning the gap of a superconducting
  flux qubit},\ }\href {https://doi.org/10.1103/PhysRevLett.102.090501}
  {\bibfield  {journal} {\bibinfo  {journal} {Phys. Rev. Lett.}\ }\textbf
  {\bibinfo {volume} {102}},\ \bibinfo {pages} {090501} (\bibinfo {year}
  {2009})}\BibitemShut {NoStop}%
\bibitem [{\citenamefont {Kalacheva}\ \emph {et~al.}(2023)\citenamefont
  {Kalacheva}, \citenamefont {Fedorov}, \citenamefont {Khrapach},\ and\
  \citenamefont {Astafiev}}]{kalacheva2023}%
  \BibitemOpen
  \bibfield  {author} {\bibinfo {author} {\bibfnamefont {D.}~\bibnamefont
  {Kalacheva}}, \bibinfo {author} {\bibfnamefont {G.}~\bibnamefont {Fedorov}},
  \bibinfo {author} {\bibfnamefont {I.}~\bibnamefont {Khrapach}},\ and\
  \bibinfo {author} {\bibfnamefont {O.}~\bibnamefont {Astafiev}},\ }\bibfield
  {title} {\bibinfo {title} {Coplanar superconducting resonators with nonlinear
  kinetic inductance bridge},\ }\href
  {https://doi.org/10.1088/1361-6668/acbc53} {\bibfield  {journal} {\bibinfo
  {journal} {Superconductor Science and Technology}\ }\textbf {\bibinfo
  {volume} {36}},\ \bibinfo {pages} {055011} (\bibinfo {year}
  {2023})}\BibitemShut {NoStop}%
\bibitem [{\citenamefont {Likharev}\ and\ \citenamefont
  {Zorin}(1985)}]{likharev1985}%
  \BibitemOpen
  \bibfield  {author} {\bibinfo {author} {\bibfnamefont {K.~K.}\ \bibnamefont
  {Likharev}}\ and\ \bibinfo {author} {\bibfnamefont {A.~B.}\ \bibnamefont
  {Zorin}},\ }\bibfield  {title} {\bibinfo {title} {Theory of the bloch-wave
  oscillations in small josephson junctions},\ }\href
  {https://doi.org/10.1007/BF00683782} {\bibfield  {journal} {\bibinfo
  {journal} {Journal of Low Temperature Physics}\ }\textbf {\bibinfo {volume}
  {59}},\ \bibinfo {pages} {347} (\bibinfo {year} {1985})}\BibitemShut
  {NoStop}%
\bibitem [{\citenamefont {Devoret}(2021)}]{devoret2021does}%
  \BibitemOpen
  \bibfield  {author} {\bibinfo {author} {\bibfnamefont {M.~H.}\ \bibnamefont
  {Devoret}},\ }\bibfield  {title} {\bibinfo {title} {Does brian josephson’s
  gauge-invariant phase difference live on a line or a circle?},\ }\href@noop
  {} {\bibfield  {journal} {\bibinfo  {journal} {Journal of Superconductivity
  and Novel Magnetism}\ }\textbf {\bibinfo {volume} {34}},\ \bibinfo {pages}
  {1633} (\bibinfo {year} {2021})}\BibitemShut {NoStop}%
\bibitem [{\citenamefont {Fedorov}\ and\ \citenamefont
  {Ustinov}(2019)}]{fedorov2019}%
  \BibitemOpen
  \bibfield  {author} {\bibinfo {author} {\bibfnamefont {G.}~\bibnamefont
  {Fedorov}}\ and\ \bibinfo {author} {\bibfnamefont {A.}~\bibnamefont
  {Ustinov}},\ }\bibfield  {title} {\bibinfo {title} {Automated analysis of
  single-tone spectroscopic data for cqed systems},\ }\href@noop {} {\bibfield
  {journal} {\bibinfo  {journal} {Quantum Science and Technology}\ }\textbf
  {\bibinfo {volume} {4}},\ \bibinfo {pages} {045009} (\bibinfo {year}
  {2019})}\BibitemShut {NoStop}%
\bibitem [{\citenamefont {Bardetski}\ and\ \citenamefont
  {Macovei}(2003)}]{bardetski2003}%
  \BibitemOpen
  \bibfield  {author} {\bibinfo {author} {\bibfnamefont {P.}~\bibnamefont
  {Bardetski}}\ and\ \bibinfo {author} {\bibfnamefont {M.}~\bibnamefont
  {Macovei}},\ }\bibfield  {title} {\bibinfo {title} {Cavity steady-state
  behaviors for a single equidistant three-level emitter},\ }\href@noop {}
  {\bibfield  {journal} {\bibinfo  {journal} {Physica Scripta}\ }\textbf
  {\bibinfo {volume} {67}},\ \bibinfo {pages} {306} (\bibinfo {year}
  {2003})}\BibitemShut {NoStop}%
\bibitem [{\citenamefont {Gasparinetti}\ \emph {et~al.}(2019)\citenamefont
  {Gasparinetti}, \citenamefont {Besse}, \citenamefont {Pechal}, \citenamefont
  {Buijs}, \citenamefont {Eichler}, \citenamefont {Carmichael},\ and\
  \citenamefont {Wallraff}}]{gasparinetti2019}%
  \BibitemOpen
  \bibfield  {author} {\bibinfo {author} {\bibfnamefont {S.}~\bibnamefont
  {Gasparinetti}}, \bibinfo {author} {\bibfnamefont {J.-C.}\ \bibnamefont
  {Besse}}, \bibinfo {author} {\bibfnamefont {M.}~\bibnamefont {Pechal}},
  \bibinfo {author} {\bibfnamefont {R.~D.}\ \bibnamefont {Buijs}}, \bibinfo
  {author} {\bibfnamefont {C.}~\bibnamefont {Eichler}}, \bibinfo {author}
  {\bibfnamefont {H.~J.}\ \bibnamefont {Carmichael}},\ and\ \bibinfo {author}
  {\bibfnamefont {A.}~\bibnamefont {Wallraff}},\ }\bibfield  {title} {\bibinfo
  {title} {Two-photon resonance fluorescence of a ladder-type atomic system},\
  }\href@noop {} {\bibfield  {journal} {\bibinfo  {journal} {Physical Review
  A}\ }\textbf {\bibinfo {volume} {100}},\ \bibinfo {pages} {033802} (\bibinfo
  {year} {2019})}\BibitemShut {NoStop}%
\bibitem [{\citenamefont {Winkel}\ \emph {et~al.}(2020)\citenamefont {Winkel},
  \citenamefont {Borisov}, \citenamefont {Gr\"unhaupt}, \citenamefont {Rieger},
  \citenamefont {Spiecker}, \citenamefont {Valenti}, \citenamefont {Ustinov},
  \citenamefont {Wernsdorfer},\ and\ \citenamefont {Pop}}]{winkel2020}%
  \BibitemOpen
  \bibfield  {author} {\bibinfo {author} {\bibfnamefont {P.}~\bibnamefont
  {Winkel}}, \bibinfo {author} {\bibfnamefont {K.}~\bibnamefont {Borisov}},
  \bibinfo {author} {\bibfnamefont {L.}~\bibnamefont {Gr\"unhaupt}}, \bibinfo
  {author} {\bibfnamefont {D.}~\bibnamefont {Rieger}}, \bibinfo {author}
  {\bibfnamefont {M.}~\bibnamefont {Spiecker}}, \bibinfo {author}
  {\bibfnamefont {F.}~\bibnamefont {Valenti}}, \bibinfo {author} {\bibfnamefont
  {A.~V.}\ \bibnamefont {Ustinov}}, \bibinfo {author} {\bibfnamefont
  {W.}~\bibnamefont {Wernsdorfer}},\ and\ \bibinfo {author} {\bibfnamefont
  {I.~M.}\ \bibnamefont {Pop}},\ }\bibfield  {title} {\bibinfo {title}
  {Implementation of a transmon qubit using superconducting granular
  aluminum},\ }\href {https://doi.org/10.1103/PhysRevX.10.031032} {\bibfield
  {journal} {\bibinfo  {journal} {Phys. Rev. X}\ }\textbf {\bibinfo {volume}
  {10}},\ \bibinfo {pages} {031032} (\bibinfo {year} {2020})}\BibitemShut
  {NoStop}%
\bibitem [{\citenamefont {Rieger}\ \emph {et~al.}(2023)\citenamefont {Rieger},
  \citenamefont {G{\"u}nzler}, \citenamefont {Spiecker}, \citenamefont
  {Paluch}, \citenamefont {Winkel}, \citenamefont {Hahn}, \citenamefont
  {Hohmann}, \citenamefont {Bacher}, \citenamefont {Wernsdorfer},\ and\
  \citenamefont {Pop}}]{rieger2023}%
  \BibitemOpen
  \bibfield  {author} {\bibinfo {author} {\bibfnamefont {D.}~\bibnamefont
  {Rieger}}, \bibinfo {author} {\bibfnamefont {S.}~\bibnamefont {G{\"u}nzler}},
  \bibinfo {author} {\bibfnamefont {M.}~\bibnamefont {Spiecker}}, \bibinfo
  {author} {\bibfnamefont {P.}~\bibnamefont {Paluch}}, \bibinfo {author}
  {\bibfnamefont {P.}~\bibnamefont {Winkel}}, \bibinfo {author} {\bibfnamefont
  {L.}~\bibnamefont {Hahn}}, \bibinfo {author} {\bibfnamefont {J.}~\bibnamefont
  {Hohmann}}, \bibinfo {author} {\bibfnamefont {A.}~\bibnamefont {Bacher}},
  \bibinfo {author} {\bibfnamefont {W.}~\bibnamefont {Wernsdorfer}},\ and\
  \bibinfo {author} {\bibfnamefont {I.}~\bibnamefont {Pop}},\ }\bibfield
  {title} {\bibinfo {title} {Granular aluminium nanojunction fluxonium qubit},\
  }\href@noop {} {\bibfield  {journal} {\bibinfo  {journal} {Nature Materials}\
  }\textbf {\bibinfo {volume} {22}},\ \bibinfo {pages} {194} (\bibinfo {year}
  {2023})}\BibitemShut {NoStop}%
\bibitem [{\citenamefont {Stehlik}\ \emph {et~al.}(2021)\citenamefont
  {Stehlik}, \citenamefont {Zajac}, \citenamefont {Underwood}, \citenamefont
  {Phung}, \citenamefont {Blair}, \citenamefont {Carnevale}, \citenamefont
  {Klaus}, \citenamefont {Keefe}, \citenamefont {Carniol}, \citenamefont
  {Kumph}, \citenamefont {Steffen},\ and\ \citenamefont {Dial}}]{stehlik2021}%
  \BibitemOpen
  \bibfield  {author} {\bibinfo {author} {\bibfnamefont {J.}~\bibnamefont
  {Stehlik}}, \bibinfo {author} {\bibfnamefont {D.~M.}\ \bibnamefont {Zajac}},
  \bibinfo {author} {\bibfnamefont {D.~L.}\ \bibnamefont {Underwood}}, \bibinfo
  {author} {\bibfnamefont {T.}~\bibnamefont {Phung}}, \bibinfo {author}
  {\bibfnamefont {J.}~\bibnamefont {Blair}}, \bibinfo {author} {\bibfnamefont
  {S.}~\bibnamefont {Carnevale}}, \bibinfo {author} {\bibfnamefont
  {D.}~\bibnamefont {Klaus}}, \bibinfo {author} {\bibfnamefont {G.~A.}\
  \bibnamefont {Keefe}}, \bibinfo {author} {\bibfnamefont {A.}~\bibnamefont
  {Carniol}}, \bibinfo {author} {\bibfnamefont {M.}~\bibnamefont {Kumph}},
  \bibinfo {author} {\bibfnamefont {M.}~\bibnamefont {Steffen}},\ and\ \bibinfo
  {author} {\bibfnamefont {O.~E.}\ \bibnamefont {Dial}},\ }\bibfield  {title}
  {\bibinfo {title} {Tunable coupling architecture for fixed-frequency transmon
  superconducting qubits},\ }\href
  {https://doi.org/10.1103/PhysRevLett.127.080505} {\bibfield  {journal}
  {\bibinfo  {journal} {Phys. Rev. Lett.}\ }\textbf {\bibinfo {volume} {127}},\
  \bibinfo {pages} {080505} (\bibinfo {year} {2021})}\BibitemShut {NoStop}%
\bibitem [{\citenamefont {{Google Quantum AI}}(2023)}]{googleAI2023}%
  \BibitemOpen
  \bibfield  {author} {\bibinfo {author} {\bibnamefont {{Google Quantum AI}}},\
  }\bibfield  {title} {\bibinfo {title} {Suppressing quantum errors by scaling
  a surface code logical qubit},\ }\href
  {https://doi.org/10.1038/s41586-022-05434-1} {\bibfield  {journal} {\bibinfo
  {journal} {Nature}\ }\textbf {\bibinfo {volume} {614}},\ \bibinfo {pages}
  {676} (\bibinfo {year} {2023})}\BibitemShut {NoStop}%
\bibitem [{\citenamefont {Gu}\ \emph {et~al.}(2017)\citenamefont {Gu},
  \citenamefont {Kockum}, \citenamefont {Miranowicz}, \citenamefont {Liu},\
  and\ \citenamefont {Nori}}]{gu2017}%
  \BibitemOpen
  \bibfield  {author} {\bibinfo {author} {\bibfnamefont {X.}~\bibnamefont
  {Gu}}, \bibinfo {author} {\bibfnamefont {A.~F.}\ \bibnamefont {Kockum}},
  \bibinfo {author} {\bibfnamefont {A.}~\bibnamefont {Miranowicz}}, \bibinfo
  {author} {\bibfnamefont {Y.-x.}\ \bibnamefont {Liu}},\ and\ \bibinfo {author}
  {\bibfnamefont {F.}~\bibnamefont {Nori}},\ }\bibfield  {title} {\bibinfo
  {title} {Microwave photonics with superconducting quantum circuits},\
  }\href@noop {} {\bibfield  {journal} {\bibinfo  {journal} {Physics Reports}\
  }\textbf {\bibinfo {volume} {718}},\ \bibinfo {pages} {1} (\bibinfo {year}
  {2017})}\BibitemShut {NoStop}%
\bibitem [{\citenamefont {Gasparinetti}\ \emph {et~al.}(2017)\citenamefont
  {Gasparinetti}, \citenamefont {Pechal}, \citenamefont {Besse}, \citenamefont
  {Mondal}, \citenamefont {Eichler},\ and\ \citenamefont
  {Wallraff}}]{gasparinetti2017}%
  \BibitemOpen
  \bibfield  {author} {\bibinfo {author} {\bibfnamefont {S.}~\bibnamefont
  {Gasparinetti}}, \bibinfo {author} {\bibfnamefont {M.}~\bibnamefont
  {Pechal}}, \bibinfo {author} {\bibfnamefont {J.-C.}\ \bibnamefont {Besse}},
  \bibinfo {author} {\bibfnamefont {M.}~\bibnamefont {Mondal}}, \bibinfo
  {author} {\bibfnamefont {C.}~\bibnamefont {Eichler}},\ and\ \bibinfo {author}
  {\bibfnamefont {A.}~\bibnamefont {Wallraff}},\ }\bibfield  {title} {\bibinfo
  {title} {Correlations and entanglement of microwave photons emitted in a
  cascade decay},\ }\href@noop {} {\bibfield  {journal} {\bibinfo  {journal}
  {Physical Review Letters}\ }\textbf {\bibinfo {volume} {119}},\ \bibinfo
  {pages} {140504} (\bibinfo {year} {2017})}\BibitemShut {NoStop}%
\bibitem [{\citenamefont {H{\"o}nigl-Decrinis}\ \emph
  {et~al.}(2018)\citenamefont {H{\"o}nigl-Decrinis}, \citenamefont {Antonov},
  \citenamefont {Shaikhaidarov}, \citenamefont {Antonov}, \citenamefont
  {Dmitriev},\ and\ \citenamefont {Astafiev}}]{honigl2018}%
  \BibitemOpen
  \bibfield  {author} {\bibinfo {author} {\bibfnamefont {T.}~\bibnamefont
  {H{\"o}nigl-Decrinis}}, \bibinfo {author} {\bibfnamefont {I.~V.}\
  \bibnamefont {Antonov}}, \bibinfo {author} {\bibfnamefont {R.}~\bibnamefont
  {Shaikhaidarov}}, \bibinfo {author} {\bibfnamefont {V.~N.}\ \bibnamefont
  {Antonov}}, \bibinfo {author} {\bibfnamefont {A.~Y.}\ \bibnamefont
  {Dmitriev}},\ and\ \bibinfo {author} {\bibfnamefont {O.~V.}\ \bibnamefont
  {Astafiev}},\ }\bibfield  {title} {\bibinfo {title} {Mixing of coherent waves
  in a single three-level artificial atom},\ }\href@noop {} {\bibfield
  {journal} {\bibinfo  {journal} {Physical Review A}\ }\textbf {\bibinfo
  {volume} {98}},\ \bibinfo {pages} {041801} (\bibinfo {year}
  {2018})}\BibitemShut {NoStop}%
\bibitem [{\citenamefont {Yan}\ \emph {et~al.}(2018)\citenamefont {Yan},
  \citenamefont {Krantz}, \citenamefont {Sung}, \citenamefont {Kjaergaard},
  \citenamefont {Campbell}, \citenamefont {Orlando}, \citenamefont
  {Gustavsson},\ and\ \citenamefont {Oliver}}]{yan2018}%
  \BibitemOpen
  \bibfield  {author} {\bibinfo {author} {\bibfnamefont {F.}~\bibnamefont
  {Yan}}, \bibinfo {author} {\bibfnamefont {P.}~\bibnamefont {Krantz}},
  \bibinfo {author} {\bibfnamefont {Y.}~\bibnamefont {Sung}}, \bibinfo {author}
  {\bibfnamefont {M.}~\bibnamefont {Kjaergaard}}, \bibinfo {author}
  {\bibfnamefont {D.~L.}\ \bibnamefont {Campbell}}, \bibinfo {author}
  {\bibfnamefont {T.~P.}\ \bibnamefont {Orlando}}, \bibinfo {author}
  {\bibfnamefont {S.}~\bibnamefont {Gustavsson}},\ and\ \bibinfo {author}
  {\bibfnamefont {W.~D.}\ \bibnamefont {Oliver}},\ }\bibfield  {title}
  {\bibinfo {title} {Tunable coupling scheme for implementing high-fidelity
  two-qubit gates},\ }\href {https://doi.org/10.1103/PhysRevApplied.10.054062}
  {\bibfield  {journal} {\bibinfo  {journal} {Phys. Rev. Appl.}\ }\textbf
  {\bibinfo {volume} {10}},\ \bibinfo {pages} {054062} (\bibinfo {year}
  {2018})}\BibitemShut {NoStop}%
\bibitem [{\citenamefont {Zhao}\ \emph {et~al.}(2020)\citenamefont {Zhao},
  \citenamefont {Xu}, \citenamefont {Lan}, \citenamefont {Chu}, \citenamefont
  {Tan}, \citenamefont {Yu},\ and\ \citenamefont {Yu}}]{zhao2020}%
  \BibitemOpen
  \bibfield  {author} {\bibinfo {author} {\bibfnamefont {P.}~\bibnamefont
  {Zhao}}, \bibinfo {author} {\bibfnamefont {P.}~\bibnamefont {Xu}}, \bibinfo
  {author} {\bibfnamefont {D.}~\bibnamefont {Lan}}, \bibinfo {author}
  {\bibfnamefont {J.}~\bibnamefont {Chu}}, \bibinfo {author} {\bibfnamefont
  {X.}~\bibnamefont {Tan}}, \bibinfo {author} {\bibfnamefont {H.}~\bibnamefont
  {Yu}},\ and\ \bibinfo {author} {\bibfnamefont {Y.}~\bibnamefont {Yu}},\
  }\bibfield  {title} {\bibinfo {title} {High-contrast $zz$ interaction using
  superconducting qubits with opposite-sign anharmonicity},\ }\href
  {https://doi.org/10.1103/PhysRevLett.125.200503} {\bibfield  {journal}
  {\bibinfo  {journal} {Phys. Rev. Lett.}\ }\textbf {\bibinfo {volume} {125}},\
  \bibinfo {pages} {200503} (\bibinfo {year} {2020})}\BibitemShut {NoStop}%
\bibitem [{\citenamefont {Fedorov}\ \emph {et~al.}(2021)\citenamefont
  {Fedorov}, \citenamefont {Remizov}, \citenamefont {Shapiro}, \citenamefont
  {Pogosov}, \citenamefont {Egorova}, \citenamefont {Tsitsilin}, \citenamefont
  {Andronik}, \citenamefont {Dobronosova}, \citenamefont {Rodionov},
  \citenamefont {Astafiev} \emph {et~al.}}]{fedorov2021}%
  \BibitemOpen
  \bibfield  {author} {\bibinfo {author} {\bibfnamefont {G.}~\bibnamefont
  {Fedorov}}, \bibinfo {author} {\bibfnamefont {S.}~\bibnamefont {Remizov}},
  \bibinfo {author} {\bibfnamefont {D.}~\bibnamefont {Shapiro}}, \bibinfo
  {author} {\bibfnamefont {W.}~\bibnamefont {Pogosov}}, \bibinfo {author}
  {\bibfnamefont {E.}~\bibnamefont {Egorova}}, \bibinfo {author} {\bibfnamefont
  {I.}~\bibnamefont {Tsitsilin}}, \bibinfo {author} {\bibfnamefont
  {M.}~\bibnamefont {Andronik}}, \bibinfo {author} {\bibfnamefont
  {A.}~\bibnamefont {Dobronosova}}, \bibinfo {author} {\bibfnamefont
  {I.}~\bibnamefont {Rodionov}}, \bibinfo {author} {\bibfnamefont
  {O.}~\bibnamefont {Astafiev}}, \emph {et~al.},\ }\bibfield  {title} {\bibinfo
  {title} {Photon transport in a bose-hubbard chain of superconducting
  artificial atoms},\ }\href@noop {} {\bibfield  {journal} {\bibinfo  {journal}
  {Physical Review Letters}\ }\textbf {\bibinfo {volume} {126}},\ \bibinfo
  {pages} {180503} (\bibinfo {year} {2021})}\BibitemShut {NoStop}%
\bibitem [{\citenamefont {Zhang}\ \emph {et~al.}(2023)\citenamefont {Zhang},
  \citenamefont {Kim}, \citenamefont {Mark}, \citenamefont {Choi},\ and\
  \citenamefont {Painter}}]{zhang2023}%
  \BibitemOpen
  \bibfield  {author} {\bibinfo {author} {\bibfnamefont {X.}~\bibnamefont
  {Zhang}}, \bibinfo {author} {\bibfnamefont {E.}~\bibnamefont {Kim}}, \bibinfo
  {author} {\bibfnamefont {D.~K.}\ \bibnamefont {Mark}}, \bibinfo {author}
  {\bibfnamefont {S.}~\bibnamefont {Choi}},\ and\ \bibinfo {author}
  {\bibfnamefont {O.}~\bibnamefont {Painter}},\ }\bibfield  {title} {\bibinfo
  {title} {A superconducting quantum simulator based on a photonic-bandgap
  metamaterial},\ }\href@noop {} {\bibfield  {journal} {\bibinfo  {journal}
  {Science}\ }\textbf {\bibinfo {volume} {379}},\ \bibinfo {pages} {278}
  (\bibinfo {year} {2023})}\BibitemShut {NoStop}%
\bibitem [{kla()}]{klayoutSh}%
  \BibitemOpen
  \href@noop {} {}\bibinfo {howpublished}
  {\url{https://github.com/shamil777/KLayout-python}}\BibitemShut {NoStop}%
\bibitem [{\citenamefont {Bruno}\ \emph {et~al.}(2015)\citenamefont {Bruno},
  \citenamefont {de~Lange}, \citenamefont {Asaad}, \citenamefont {van~der
  Enden}, \citenamefont {Langford},\ and\ \citenamefont {DiCarlo}}]{bruno2015}%
  \BibitemOpen
  \bibfield  {author} {\bibinfo {author} {\bibfnamefont {A.}~\bibnamefont
  {Bruno}}, \bibinfo {author} {\bibfnamefont {G.}~\bibnamefont {de~Lange}},
  \bibinfo {author} {\bibfnamefont {S.}~\bibnamefont {Asaad}}, \bibinfo
  {author} {\bibfnamefont {K.~L.}\ \bibnamefont {van~der Enden}}, \bibinfo
  {author} {\bibfnamefont {N.~K.}\ \bibnamefont {Langford}},\ and\ \bibinfo
  {author} {\bibfnamefont {L.}~\bibnamefont {DiCarlo}},\ }\bibfield  {title}
  {\bibinfo {title} {{Reducing intrinsic loss in superconducting resonators by
  surface treatment and deep etching of silicon substrates}},\ }\bibfield
  {journal} {\bibinfo  {journal} {Applied Physics Letters}\ }\textbf {\bibinfo
  {volume} {106}},\ \href {https://doi.org/10.1063/1.4919761}
  {10.1063/1.4919761} (\bibinfo {year} {2015}),\ \bibinfo {note}
  {182601}\BibitemShut {NoStop}%
\bibitem [{\citenamefont {Kalacheva}\ \emph {et~al.}(2020)\citenamefont
  {Kalacheva}, \citenamefont {Fedorov}, \citenamefont {Kulakova}, \citenamefont
  {Zotova}, \citenamefont {Korostylev}, \citenamefont {Khrapach}, \citenamefont
  {Ustinov},\ and\ \citenamefont {Astafiev}}]{kalacheva2020}%
  \BibitemOpen
  \bibfield  {author} {\bibinfo {author} {\bibfnamefont {D.}~\bibnamefont
  {Kalacheva}}, \bibinfo {author} {\bibfnamefont {G.}~\bibnamefont {Fedorov}},
  \bibinfo {author} {\bibfnamefont {A.}~\bibnamefont {Kulakova}}, \bibinfo
  {author} {\bibfnamefont {J.}~\bibnamefont {Zotova}}, \bibinfo {author}
  {\bibfnamefont {E.}~\bibnamefont {Korostylev}}, \bibinfo {author}
  {\bibfnamefont {I.}~\bibnamefont {Khrapach}}, \bibinfo {author}
  {\bibfnamefont {A.~V.}\ \bibnamefont {Ustinov}},\ and\ \bibinfo {author}
  {\bibfnamefont {O.~V.}\ \bibnamefont {Astafiev}},\ }\bibfield  {title}
  {\bibinfo {title} {Improving the quality factor of superconducting resonators
  by post-process surface treatment},\ }\href
  {https://doi.org/10.1063/5.0011900} {\bibfield  {journal} {\bibinfo
  {journal} {AIP Conference Proceedings}\ }\textbf {\bibinfo {volume} {2241}},\
  \bibinfo {pages} {020018} (\bibinfo {year} {2020})}\BibitemShut {NoStop}%
\bibitem [{\citenamefont {Liechao}\ \emph {et~al.}(2012)\citenamefont
  {Liechao}, \citenamefont {Kang},\ and\ \citenamefont {Wen}}]{liechao2012}%
  \BibitemOpen
  \bibfield  {author} {\bibinfo {author} {\bibfnamefont {L.}~\bibnamefont
  {Liechao}}, \bibinfo {author} {\bibfnamefont {J.}~\bibnamefont {Kang}},\ and\
  \bibinfo {author} {\bibfnamefont {J.}~\bibnamefont {Wen}},\ }\bibfield
  {title} {\bibinfo {title} {Low stress tin as metal hard mask for advance
  cu-interconnect},\ }\href {https://doi.org/10.1149/1.3694357} {\bibfield
  {journal} {\bibinfo  {journal} {ECS Transactions}\ }\textbf {\bibinfo
  {volume} {44}},\ \bibinfo {pages} {481} (\bibinfo {year} {2012})}\BibitemShut
  {NoStop}%
\bibitem [{\citenamefont {Dolan}(1977)}]{dolan1977}%
  \BibitemOpen
  \bibfield  {author} {\bibinfo {author} {\bibfnamefont {G.~J.}\ \bibnamefont
  {Dolan}},\ }\bibfield  {title} {\bibinfo {title} {Offset masks for lift-off
  photoprocessing},\ }\href {https://aip.scitation.org/doi/10.1063/1.89690}
  {\bibfield  {journal} {\bibinfo  {journal} {Appl. Phys. Lett.}\ }\textbf
  {\bibinfo {volume} {31}},\ \bibinfo {pages} {337} (\bibinfo {year}
  {1977})}\BibitemShut {NoStop}%
\bibitem [{\citenamefont {Osman}\ \emph {et~al.}(2021)\citenamefont {Osman},
  \citenamefont {Simon}, \citenamefont {Bengtsson}, \citenamefont {Kosen},
  \citenamefont {Krantz}, \citenamefont {P.~Lozano}, \citenamefont
  {Scigliuzzo}, \citenamefont {Delsing}, \citenamefont {Bylander},\ and\
  \citenamefont {Fadavi~Roudsari}}]{osman2021}%
  \BibitemOpen
  \bibfield  {author} {\bibinfo {author} {\bibfnamefont {A.}~\bibnamefont
  {Osman}}, \bibinfo {author} {\bibfnamefont {J.}~\bibnamefont {Simon}},
  \bibinfo {author} {\bibfnamefont {A.}~\bibnamefont {Bengtsson}}, \bibinfo
  {author} {\bibfnamefont {S.}~\bibnamefont {Kosen}}, \bibinfo {author}
  {\bibfnamefont {P.}~\bibnamefont {Krantz}}, \bibinfo {author} {\bibfnamefont
  {D.}~\bibnamefont {P.~Lozano}}, \bibinfo {author} {\bibfnamefont
  {M.}~\bibnamefont {Scigliuzzo}}, \bibinfo {author} {\bibfnamefont
  {P.}~\bibnamefont {Delsing}}, \bibinfo {author} {\bibfnamefont
  {J.}~\bibnamefont {Bylander}},\ and\ \bibinfo {author} {\bibfnamefont
  {A.}~\bibnamefont {Fadavi~Roudsari}},\ }\bibfield  {title} {\bibinfo {title}
  {Simplified josephson-junction fabrication process for reproducibly
  high-performance superconducting qubits},\ }\href
  {https://doi.org/10.1063/5.0037093} {\bibfield  {journal} {\bibinfo
  {journal} {Applied Physics Letters}\ }\textbf {\bibinfo {volume} {118}},\
  \bibinfo {pages} {064002} (\bibinfo {year} {2021})},\ \Eprint
  {https://arxiv.org/abs/https://doi.org/10.1063/5.0037093}
  {https://doi.org/10.1063/5.0037093} \BibitemShut {NoStop}%
\bibitem [{\citenamefont {Chen}\ \emph {et~al.}(2014)\citenamefont {Chen},
  \citenamefont {Megrant}, \citenamefont {Kelly}, \citenamefont {Barends},
  \citenamefont {Bochmann}, \citenamefont {Chen}, \citenamefont {Chiaro},
  \citenamefont {Dunsworth}, \citenamefont {Jeffrey}, \citenamefont {Mutus},
  \citenamefont {O'Malley}, \citenamefont {Neill}, \citenamefont {Roushan},
  \citenamefont {Sank}, \citenamefont {Vainsencher}, \citenamefont {Wenner},
  \citenamefont {White}, \citenamefont {Cleland},\ and\ \citenamefont
  {Martinis}}]{chen2014}%
  \BibitemOpen
  \bibfield  {author} {\bibinfo {author} {\bibfnamefont {Z.}~\bibnamefont
  {Chen}}, \bibinfo {author} {\bibfnamefont {A.}~\bibnamefont {Megrant}},
  \bibinfo {author} {\bibfnamefont {J.}~\bibnamefont {Kelly}}, \bibinfo
  {author} {\bibfnamefont {R.}~\bibnamefont {Barends}}, \bibinfo {author}
  {\bibfnamefont {J.}~\bibnamefont {Bochmann}}, \bibinfo {author}
  {\bibfnamefont {Y.}~\bibnamefont {Chen}}, \bibinfo {author} {\bibfnamefont
  {B.}~\bibnamefont {Chiaro}}, \bibinfo {author} {\bibfnamefont
  {A.}~\bibnamefont {Dunsworth}}, \bibinfo {author} {\bibfnamefont
  {E.}~\bibnamefont {Jeffrey}}, \bibinfo {author} {\bibfnamefont {J.~Y.}\
  \bibnamefont {Mutus}}, \bibinfo {author} {\bibfnamefont {P.~J.~J.}\
  \bibnamefont {O'Malley}}, \bibinfo {author} {\bibfnamefont {C.}~\bibnamefont
  {Neill}}, \bibinfo {author} {\bibfnamefont {P.}~\bibnamefont {Roushan}},
  \bibinfo {author} {\bibfnamefont {D.}~\bibnamefont {Sank}}, \bibinfo {author}
  {\bibfnamefont {A.}~\bibnamefont {Vainsencher}}, \bibinfo {author}
  {\bibfnamefont {J.}~\bibnamefont {Wenner}}, \bibinfo {author} {\bibfnamefont
  {T.~C.}\ \bibnamefont {White}}, \bibinfo {author} {\bibfnamefont {A.~N.}\
  \bibnamefont {Cleland}},\ and\ \bibinfo {author} {\bibfnamefont {J.~M.}\
  \bibnamefont {Martinis}},\ }\bibfield  {title} {\bibinfo {title} {Fabrication
  and characterization of aluminum airbridges for superconducting microwave
  circuits},\ }\href {https://doi.org/10.1063/1.4863745} {\bibfield  {journal}
  {\bibinfo  {journal} {Applied Physics Letters}\ }\textbf {\bibinfo {volume}
  {104}},\ \bibinfo {pages} {052602} (\bibinfo {year} {2014})},\ \Eprint
  {https://arxiv.org/abs/https://doi.org/10.1063/1.4863745}
  {https://doi.org/10.1063/1.4863745} \BibitemShut {NoStop}%
\bibitem [{\citenamefont {Schuster}\ \emph {et~al.}(2005)\citenamefont
  {Schuster}, \citenamefont {Wallraff}, \citenamefont {Blais}, \citenamefont
  {Frunzio}, \citenamefont {Huang}, \citenamefont {Majer}, \citenamefont
  {Girvin},\ and\ \citenamefont {Schoelkopf}}]{schuster2005}%
  \BibitemOpen
  \bibfield  {author} {\bibinfo {author} {\bibfnamefont {D.~I.}\ \bibnamefont
  {Schuster}}, \bibinfo {author} {\bibfnamefont {A.}~\bibnamefont {Wallraff}},
  \bibinfo {author} {\bibfnamefont {A.}~\bibnamefont {Blais}}, \bibinfo
  {author} {\bibfnamefont {L.}~\bibnamefont {Frunzio}}, \bibinfo {author}
  {\bibfnamefont {R.-S.}\ \bibnamefont {Huang}}, \bibinfo {author}
  {\bibfnamefont {J.}~\bibnamefont {Majer}}, \bibinfo {author} {\bibfnamefont
  {S.~M.}\ \bibnamefont {Girvin}},\ and\ \bibinfo {author} {\bibfnamefont
  {R.~J.}\ \bibnamefont {Schoelkopf}},\ }\bibfield  {title} {\bibinfo {title}
  {ac stark shift and dephasing of a superconducting qubit strongly coupled to
  a cavity field},\ }\href {https://doi.org/10.1103/PhysRevLett.94.123602}
  {\bibfield  {journal} {\bibinfo  {journal} {Phys. Rev. Lett.}\ }\textbf
  {\bibinfo {volume} {94}},\ \bibinfo {pages} {123602} (\bibinfo {year}
  {2005})}\BibitemShut {NoStop}%
\end{thebibliography}%

\end{document}